\documentclass{jpp}
\usepackage{graphicx}

\usepackage[utf8]{inputenc}
\usepackage[T1]{fontenc}
\usepackage{amsmath}

\usepackage{xcolor}
\definecolor{newcolor}{rgb}{.8,.349,.1}

\usepackage{siunitx}

\usepackage{tikz}
\usetikzlibrary{shapes,arrows}

\newcommand{\ch}{{\rm ch}}
\renewcommand{\sh}{{\rm sh}}

\newcommand{\ex}{\mathbf{e}_{\rm x}}
\newcommand{\ey}{\mathbf{e}_{\rm y}}
\newcommand{\ez}{\mathbf{e}_{\rm z}}
\newcommand{\ephi}{\mathbf{e}_{\varphi}}

\newcommand{\rLarmor}{r_{\rm L}}

\usepackage{numprint}

\usepackage{amsmath, amssymb}

\usepackage[linesnumbered,ruled,vlined]{algorithm2e}
\SetKwRepeat{Do}{do}{while}%

\SetCommentSty{mycommfont}

\SetKwInput{KwInput}{Input}                
\SetKwInput{KwOutput}{Output}              

\shorttitle{Relativistic particle mover}
\shortauthor{J. P\'etri}

\title[A relativistic particle pusher for ultra-strong fields]{A relativistic particle pusher for ultra-strong electromagnetic fields}

\author{J. P\'etri\aff{1}
	\corresp{\email{jerome.petri@astro.unistra.fr}}}

\affiliation{\aff{1}Universit\'e de Strasbourg, CNRS, Observatoire astronomique de Strasbourg, UMR 7550, F-67000 Strasbourg, France.}

\begin{document}
	
\maketitle

\begin{abstract}
Kinetic plasma simulations are nowadays commonly used to study a wealth of non-linear behaviours and properties in laboratory and space plasmas. In particular, in high-energy physics and astrophysics, the plasma usually evolves in ultra-strong electromagnetic fields produced by intense laser beams for the former or by rotating compact objects such as neutron stars and black holes for the latter. In these ultra-strong electromagnetic fields, the gyro-period is several orders of magnitude smaller than the timescale on which we desire to investigate the plasma evolution. Some approximations are required like for instance artificially decreasing the electromagnetic field strength which is certainly not satisfactory. The main flaw of this downscaling is that it cannot reproduce particle acceleration to ultra-relativistic speeds with Lorentz factor above $\gamma \approx 10^3-10^4$. In this paper, we design a new algorithm able to catch particle motion and acceleration to Lorentz factor up to $10^{15}$ or even higher by using Lorentz boosts to special frames where the electric and magnetic field are parallel. Assuming that these fields are locally uniform in space and constant in time, we solve analytically the equation of motion in a tiny region smaller than the length scale of the spatial and temporal gradient of the field. This analytical integration of the orbit severely reduces the constrain on the time step, allowing us to use large time steps, avoiding to resolved the ultra high gyro-frequency. We performed simulations in ultra-strong spatially and time dependent electromagnetic fields, showing that our particle pusher is able to follow accurately the exact analytical solution for very long times. This property is crucial to properly capture for instance lepton electrodynamics in electromagnetic waves produced by fast rotating neutron stars. We conclude with a simple implementation our new pusher into a one dimensional relativistic electromagnetic particle-in-cell code, testing it against plasma oscillations, two-stream instabilities and strongly magnetized relativistic shocks.
\end{abstract}

\section{Introduction}

With the advent of numerical simulation techniques and the increasing computational power capabilities, plasma physics has benefited from a better and deeper description of its behaviour and properties in many contexts of laboratory experiments and space plasmas. Nowadays numerical simulations play a key role in the development of our knowledges about plasmas. However, these kinetic simulations still suffers from limitations due to the current hardware capabilities. For instance, the Vlasov-Maxwell equations require 3~dimensions in physical space as well as 3~dimensions in velocity space thus a total 6~dimensions, forbidding us to perform high resolution simulations. Another limitation comes from the different time and space scales to be resolved in order to properly catch the collective plasma effects. The simulation box is usually much larger than the gyro-radius or than the skin depth. Another restriction of particular interest in high-energy physics and astrophysics is the motion of plasmas in ultra-strong electromagnetic fields. By ultra-strong we mean field strengths about the quantum critical value of $B \approx \numprint{4.4e9}$~\si{\tesla}. Such fields are commonly met in neutron star magnetospheres like pulsars \citep{petri_theory_2016} and magnetars \citep{mereghetti_magnetars:_2015}. For pulsars, the magnetic field strength is typically $B\approx \numprint{e8}$~\si{\tesla} whereas for magnetars it easily exceeds $B\approx \numprint{e10}$~\si{\tesla}. For comparison, the dynamical time scales of interest are of the order of seconds and hours, many orders of magnitude longer than the gyro-period in such fields. These values put stringent constrains on the time step of any numerical algorithm because the gyration period is orders of magnitude smaller than the macroscopic evolution period given by the neutron star rotation frequency~$\Omega=2\,\pi/P$ with $P$ its period of revolution. Indeed, the ratio between the Larmor frequency and the stellar rotation frequency is
\begin{equation}
\frac{\omega_B}{\Omega} = \frac{q\,B}{m\,\Omega} = \numprint{2.8e18} \, \left( \frac{P}{1~\si{\second}} \right) \, \left( \frac{B}{\numprint{e8}~\si{\tesla}} \right) .
\end{equation}
$B$ is the stellar magnetic field strength, $\Omega$ its angular velocity, $q$ the particle charge and $m$ its mass. For the above numerical applications, we assumed electrons or positrons to be the main constituents of the magnetospheric plasma.
Moreover, the Larmor radius associated to these fields is
\begin{equation}
\rLarmor = \frac{\gamma\,m\,c}{q\,B} = \numprint{1.7e-5}~\si{\meter} \, \left( \frac{\gamma}{\numprint{e6}} \right)  \, \left( \frac{B}{\numprint{e8}~\si{\tesla}} \right)^{-1} .
\end{equation}
$c$ is the speed of light and $\gamma$ the particle Lorentz factor. This length scale remains much smaller than the typical size of a neutron star estimated to be about $R_{\rm ns} = 10$~\si{\kilo\meter}. Thus the ratio between Larmor radius and neutron star radius is about $\epsilon = \rLarmor / R_{\rm ns} = \numprint{e-10}$, allowing to separate both scales. Note that $R_{\rm ns}$ is also the typical length scale for the electromagnetic field gradient. To a very good approximation, we can assert that leptons orbiting in this field feel an almost constant and uniform electromagnetic field during thousands to millions of gyro periods.

Particle orbits in a plane electromagnetic wave are described by a one parameter family solution given by the so called strength parameter defined by
\begin{equation}\label{eq:StrengthParameter}
a = \frac{q\,E}{m\,c\,\omega}
\end{equation}
$m$ and $q$ are the particle mass and charge respectively, $E$ is the amplitude of the wave electric field and $\omega$ its frequency. This strength parameter is ridiculously small for visible light with $a_{\rm light} \approx 10^{-10}$, but substantial for high intensity laser with $a_{\rm laser} \approx 10^2$ and dramatically high at the surface of neutron stars with $a_{\rm ns} \approx 10^{18}$.

Many kinetic codes have been designed to solve the pulsar magnetosphere problem. Unfortunately, none of the PIC results presented so far have been able to put realistic electromagnetic field strengths into the simulation box \citep{belyaev_dissipation_2015, philippov_ab_2014, cerutti_particle_2015}. This questions the veracity of those works that even add radiation reaction in a regime not corresponding to what is expected in neutron stars \citep{cerutti_modelling_2016}. Some scaling technique is proposed to extrapolate simulation results to realistic values \citep{kalapotharakos_three-dimensional_2018} but such assertions must be checked by direct numerical computations in physically self-consistent fields.

Recently, \cite{zenitani_boris_2018} improved the standard \cite{boris_relativistic_1970} particle pusher by computing the exact analytical rotation in the magnetic part of the Lorentz force. They showed an improvement in the accuracy. The Boris algorithm is popular because it is simple and accurate. Its stability is accounted for by its phase space volume preserving properties as shown by \cite{qin_why_2013}. \cite{umeda_three-step_2018} proposed an improvement of the Boris algorithm by employing a three stage step. Relativistic simulations are even more stringent about numerical error accumulation and volume preserving schemes are highly recommended as pointed out by \cite{zhang_volume-preserving_2015}. However, these recent papers did not address the problem of particle pusher in very strong electromagnetic fields as those existing in neutron stars.

Efficient particle pushers in ultra-strong electromagnetic fields are  nevertheless a fundamental prerequisite to simulate the true electrodynamics in neutron star magnetospheres such as pulsars and magnetars. Specific algorithms have been designed for motion in strong magnetic fields to solve for kinetic plasma problems via numerical simulations. For instance a Vlasov-Poisson approach was tackled by \cite{crouseilles_uniformly_2017} to solve motion in a strong and uniform external magnetic field using a two-scale formalism. The high frequency gyration about the magnetic field is decoupled from the secular evolution occurring on a much longer time scale. Semi-implicit scheme in velocity space preserving the asymptotic limit of the guiding centre approximation were also investigated by \cite{filbet_asymptotically_2015}. \cite{geiser_integrators_2016} discussed the merit of several integrators used in electromagnetic PIC simulations, like the explicit and implicit Boris algorithm and a cyclotronic integrator. An explicit time-reversible cyclotronic integrator has been derived in \cite{patacchini_explicit_2009} to avoid a fine resolution of the Larmor frequency on some field configurations. Velocity Verlet algorithms \citep{verlet_computer_1967} for strong homogeneous static external magnetic fields in the context of molecular dynamics have been investigated by \cite{spreiter_classical_1999} performing a Taylor expansion. For a comprehensive comparison of relativistic particle integrators, see \cite{ripperda_comprehensive_2018}. These authors carefully compared the merit of the standard Boris algorithm \cite{boris_relativistic_1970}, the \cite{vay_simulation_2008} implicit scheme in space velocity, the \cite{higuera_structure-preserving_2017} second order method and the implicit midpoint method described in \cite{lapenta_particle_2011}. This leapfrog scheme already appeared in \cite{verboncoeur_particle_2005}. A fully implicit update in space and velocity parameters for relativistic particle integrators relying on \cite{vay_simulation_2008} velocity advance has been explored by \cite{petri_fully_2017}. There it has been shown that catching properly and accurately the simple electric drift motion in an ultra relativistic regime remains extremely difficult to achieve. Unfortunately, neutron star magnetospheres are common places for such relativistic drift velocities. It is therefore compulsory to design efficient and accurate numerical schemes to faithfully follow these trajectories. This is a crucial step towards realistic particle acceleration and radiation in ultra-strong electromagnetic fields. Some tests of particle acceleration in a plane electromagnetic wave using standard pushers and reported by \cite{arefiev_temporal_2015} but showing severe limitations in the accuracy already for modest strength parameters around $a \approx 20$. This is unacceptable for our investigation of neutron star magnetospheres. \cite{giovanelli_analytic_1987} showed how to use exact analytical solutions in constant electromagnetic fields to construct new particle pushers by transformations to appropriate frames where electric and magnetic fields are parallel. Our ideas follow the same line with an extensive test of the algorithm in several geometric configurations. \cite{gordon_pushing_2017} discussed the problem of the Boris algorithm for ultra-strong fields. They used an unsplit covariant pusher and showed how to include radiation reaction.

Moreover, when accelerated to very high energies in the high fields of neutron stars, particles are subject to radiation reaction, a stringent friction damping the motion by limiting the Lorentz factor to \numprint{e8} or \numprint{e9} according to current wisdom. Unfortunately, there is no unique model for implementing this damping into a numerical code. \cite{vranic_classical_2016} proposed a careful study of several radiation reaction forces to be used in classical PIC codes. They showed that all prescriptions for this force give similar results. For pulsars, radiation reaction is inescapable for several reason. First, pulsars are known to be very efficient particle accelerators, pushing electron-positron pairs to Lorentz factors as high as $ \gamma \approx \numprint{e9}$. See for instance the simulations performed in vacuum by \cite{petri_pulsar_2019} assuming radiation reaction force operates in balance with the accelerating electric field. Without radiation reaction, particles would tend to much higher Lorentz factors \citep{finkbeiner_effects_1989}. Second, pulsars are also famous for being high energy emitters, producing photons with energy above the GeV even up to TeV ranges for the Crab pulsar \citep{ansoldi_teraelectronvolt_2016} and around tenth of TeV for the Vela pulsar \citep{djannati-atai_h.e.s.s._2017}. Therefore radiation inevitably has to impact on the particle dynamics in a non perturbative way. Implementing this additional radiative force into our code is under progress. One approach would be to find exact analytical solutions for constant electromagnetic fields including radiation reaction. Some analytical solutions exist in special cases, for instance in a transverse electromagnetic wave \citep{piazza_exact_2008, hadad_effects_2010}. Solutions in homogeneous and constant electromagnetic field are also known \citep{heintzmann_exact_1973}. See also \cite{shen_radiation_1978} and \cite{laue_acceleration_1986, biltzinger_selfconsistent_2000} for applications to neutron star magnetospheres. A second approach consist to find an approximate analytical expression for the change in velocity and Lorentz factor during an integration time step for the Lorentz force solely and then correct for radiation reaction (which should remain weak compared to the Lorentz force). The third brute force approach would resolve the full motion including the radiation reaction but at the expense of requiring much smaller time steps. All these options are currently investigated and tested but will be shown in another work.

As particle simulation codes are prone to discrete particle noise scaling as $1/\sqrt{N_{\rm par}}$ where $N_{\rm par}$ is the number of particles used in the simulation box, a large number of particles is necessary to achieve low level noise results. High performance computing optimizing memory usage and CPU time is recommended \citep{bowers_ultrahigh_2008}. However, in this paper we rather focus on achieving very high accuracy in the numerical solution sticking as close as possible to the analytical solutions of the physical problem. Therefore, in this study, computational time is not an issue but setting realistic ultra-strong electromagnetic fields is a stringent and critical issue to be able to properly capture the correct physics. Actually, having not such a severe restriction on the time step compared to other algorithms, the computational extra cost is largely compensated by taking time steps that are thousands to millions or billions of time larger than those required by explicit time integrators. We also expect that parallelization techniques such as Genera Purpose Graphics Processing Units (GPGPUs) and Message Passing Interface (MPI) will speed up the extra cost of performing the analytical computations of the solutions.

In this paper, we first expose the general idea, detailing the method and the algorithm in section~\ref{sec:Method}. Then we remind the exact analytical solutions for a charged particle in an arbitrary electromagnetic field in section~\ref{sec:Motion} and show how to switch to a frame where $E$ and $B$ are parallel. The time stepping is critical for efficient implementation in any PIC code and is discussed in section~\ref{sec:TimeStep}. We then test our implementation of these equations in several configurations for which exact analytical solutions are known, see section~\ref{sec:Results}. Some simple implementations of our new method to a 1D relativistic PIC code for periodic boundary conditions is presented in section~\ref{sec:PIC}. We conclude about our achievements and possible extensions in section~\ref{sec:Conclusion}.

\section{General method and algorithm}
\label{sec:Method}

Our main goal in this paper is to use exact analytical expressions for the particle trajectories in a homogeneous and uniform electromagnetic field. As such expressions are known and very handy in a frame where the electric field~$\mathbf{E}$ and magnetic field~$\mathbf{B}$ are parallel, we introduce two important frames to perform our numerical simulations. First, we denote by $K$ the observer frame in which we want to evolve the particle motion one time step. Second, we consider a new frame $K'$ where the electric field and magnetic field are parallel. We will demonstrate that except for the case where $E=c\,B$ and $\mathbf{E} \cdot \mathbf{B}=0$ simultaneously, corresponding to a plane electromagnetic wave propagating in vacuum, there always exist one such frame. An excellent and very detailed reference one this topic is \cite{gourgoulhon_relativite_2010}, but see also \cite{sengupta_classical_2007}.

Kinematic and dynamical quantities like position and velocity are transformed according to the special relativistic Lorentz transform valid for any four vector~$\mathbb{A}$. In our notations, the temporal components are labelled with index~$0$ whereas the spatial components are labelled with indices running from $1$ to $3$. The minskowskian metric is given by the diagonal matrix $\eta_{ik}={\rm diag}(+1,-1,-1,-1)$. In particular, the contravariant components~$A^i = (A^0, \mathbf{A})$ of this four vector~$\mathbb{A}$ in both frames are related by
\begin{subequations}
	\label{eq:TransfoLorentz}
	\begin{align}
	A'^0 & = \Gamma \, ( A^0 - \mathbf{A} \cdot \mathbf{V}/c ) \\
	A'_\parallel & = \Gamma \, ( A_\parallel - V \, A_0 /c  ) \\
	\mathbf{A}'_\perp & = \mathbf{A}_\perp .
	\end{align}
\end{subequations}
$A_\parallel$ and $A'_\parallel$ are the components along the relative velocity between both frames, $A_\perp$ and $A'_\perp$ the components perpendicular to this relative velocity, $\mathbf{V}$ is the 3-velocity vector of the frame $K'$ with respect to the frame $K$ and $\Gamma = (1-V^2/c^2)^{-1/2}$ the associated Lorentz factor. In the following sections, we specialize the basis vectors such that the $z$ and $z'$ axis are aligned along $\mathbf{V}$. Note that the particle proper frame is never used, only its proper time is required to compute the trajectories.

The procedure to advance the particle position and velocity one time step is then the following. Compute both relativistic electromagnetic invariants $\mathcal{I}_1 = E^2-c^2\,B^2$ and $\mathcal{I}_2 = \mathbf{E} \cdot \mathbf{B}$. If $\mathcal{I}_2=0$ and $\mathcal{I}_1\neq0$, then a frame where either the electric field or the magnetic field vanishes exists, depending on the sign of~$\mathcal{I}_1$. We switch to this new frame $K'$ and solve analytically the equation of motion. If $\mathcal{I}_1=0$, we have to solve the motion separately as no physical frame $K'$ exist with speed strictly less than~$c$ where $\mathbf{E}$ and $\mathbf{B}$ are parallel. This special case is called a null or light like field. If on the other side $\mathcal{I}_2\neq0$, there always exists a frame $K'$ where $\mathbf{E}$ and $\mathbf{B}$ are parallel. Then if possible switch to the new frame $K'$ by a Lorentz boost. Solve the particle motion in $K'$ and then Lorentz boost back to $K$. In the frame $K'$, if the $z'$ axis is not aligned with the common direction of $\mathbf{E}$ and $\mathbf{B}$, we also apply an Euler rotation to bring the new $z''$ axis along this direction.

To summarize all the cases, we show a pseudo-code explaining how to evolve the particle trajectory depending on the field configuration in table~\ref{tab:code}. The next step requires the solution of the 4-velocity and 4-position in the frame where $\mathbf{E}$ and $\mathbf{B}$ are parallel. This is exposed in the next section.
\begin{table}
	\begin{algorithm}[H]
		\DontPrintSemicolon
		
		\KwInput{The initial 4-position and 4-velocity of the particle $(x^n,u^n)$ at time $t^n$. \\ The electromagnetic invariants $\mathcal{I}_1 = E^2-c^2\,B^2$ and $\mathcal{I}_2 = \mathbf{E} \cdot \mathbf{B}$.}
		\KwOutput{The final 4-position and 4-velocity of the particle $(x^{n+1},u^{n+1})$ at time $t^{n+1}$.}
		\KwData{Electromagnetic field $(\mathbf{E},\mathbf{B})$ at particle position~$x^n$.}
		
		\tcc{Check for zero electromagnetic field $E==B==0$}
		\If{($E==0$ \&$ B==0$)}
		{ Integrate particle trajectory according to eq.~(\ref{eq:champ_faible_position}), no update in velocity.}
		
		\tcc{Check for light-like wave $\mathcal{I}_1 == 0$ \& $\mathcal{I}_2 == 0$}
		\ElseIf{($\mathcal{I}_1 == 0$ \& $\mathcal{I}_2 == 0$)} 
		{Integrate particle trajectory according to eq.~(\ref{eq:light_like_velocity}), (\ref{eq:light_like_position}) }
		
		\tcc{Check for zero magnetic field $B==0$}
		\ElseIf{($B==0$)}
		{ Integrate particle trajectory according to eq.~(\ref{eq:no_magnetic_velocity}), (\ref{eq:no_magnetic_position}) }
		
		\tcc{Check for zero electric field $E==0$}			
		\ElseIf{($E==0$)} 
		{Integrate particle trajectory according to eq.~(\ref{eq:no_electric_velocity}), (\ref{eq:no_electric_position}) }
		
		\tcc{Otherwise integrate in frame where $\mathbf{E}$ and $\mathbf{B}$ are parallel}			
		\Else
		{
			{Integrate particle trajectory according to eq.~(\ref{eq:VitesseParticule}), (\ref{eq:TrajectoireParticule}) }
		}
		
		\caption{The algorithm to solve for particle motion.}
	\end{algorithm}
	\caption{\label{tab:code}Pseudo-code summarizing the full algorithm. It shows the many special cases to evaluate and handle (light-like fields, orthogonal fields, arbitrary fields).}
\end{table}

\section{Charge in an uniform electromagnetic field}
\label{sec:Motion}

In this section, we derive the exact analytical solution of a charged particle in relativistic motion in an uniform electromagnetic field for an arbitrary geometric configuration. To do this we first find a frame where electric and magnetic fields $\mathbf{E}$ and $\mathbf{B}$ are parallel. Next we solve exactly and analytically the equation of motion in the relativistic regime where the electric and the magnetic field are parallel. These solutions are also presented in \cite{gourgoulhon_relativite_2010} and in \cite{jackson_electrodynamique_2001} in a somewhat different way with different integration constants.

A direct integration of the equation of motion in the observer frame needs to solve for the eigenvalues and eigenvectors of the antisymmetric electromagnetic tensor~$F^{ik}$. This has been performed by \cite{vandervoort_relativistic_1960} but we found it easier to first switch to the special frame where $\mathbf{E}$ and $\mathbf{B}$ are parallel and then compute the solution. This is also the strategy we adopt in our numerical implementation of the algorithm. It is therefore also necessary to readjust the axes to conform to the orientation we employ in the subsequent paragraphs.

\subsection{Frame where $\mathbf{E}$ and $\mathbf{B}$ are parallel}

A useful way to follow particle trajectories in any prescribed electromagnetic field given in an inertial frame~$K$ consists to Lorentz transform the electromagnetic field into a special frame~$K'$ in which the electric field is parallel to the magnetic field. In the general case, there is an infinite number of frames for which the electric field is parallel to the magnetic field. We can however choose the particular velocity given by $\mathbf{V}_\parallel = c\,\boldsymbol{\beta}_\parallel = \alpha \, \mathbf E \wedge \mathbf B$ where $\alpha$ is a constant to be determined. In the frame~$K'$, the motion is along the common direction of $\mathbf{E}'$ and $\mathbf{B}'$ that is the electromagnetic field as measured in this frame~$K'$. The constant $\alpha$ is the solution given by
\begin{equation}
\alpha = \frac{E^2+c^2\,B^2 - \sqrt{\mathcal{I}_1^2 + 4\,c^2\,\mathcal{I}_2^2}}{2 \, (\mathbf{E} \wedge \mathbf{B})^2} .
\end{equation}
The minus sign in front of the square root enforces a speed less than that of light. The electric and magnetic fields in the frame moving at speed $\mathbf{V}_\parallel$ are found by a special relativistic Lorentz boost of the electromagnetic field and gives
\begin{subequations}
	\label{eq:transform_EB}
	\begin{align}
	\mathbf{E}' & = \Gamma_\parallel \, [ ( 1 - \alpha \, B^2 ) \, \mathbf{E} + \alpha \, ( \mathbf{E} \cdot \mathbf{B}) \mathbf{B} ] \\
	\mathbf{B}' & = \Gamma_\parallel \, [ ( 1 - \alpha \, E^2/c^2 ) \, \mathbf{B} + \alpha \, ( \mathbf{E} \cdot \mathbf{B}) \mathbf{E}/c^2 ] .
	\end{align}
\end{subequations}
The Lorentz factor of the frame in which $\mathbf{E}'$ and $\mathbf{B}'$ are parallel is defined by $\Gamma_\parallel = (1-\beta_\parallel^2)^{-1/2}$. In this frame, the particle trajectory is decomposed into a motion along the common direction of $\mathbf{E}'$ and $\mathbf{B}'$ and a gyration around the magnetic field $\mathbf{B}'$. Thus the local tangent to the trajectory becomes $\mathbf{t}'_\parallel = \pm \mathbf{E}'/E' = \pm \mathbf{B}'/B'$, the sign being chosen such that particles flow outwards. Our expression for the particle velocity resembles to the Aristotelian expression given by \cite{gruzinov_aristotelian_2013}. Our velocity prescription is however more general because we do not assume that particles travel exactly at the speed of light. The speed along the common $\mathbf{E}$ and $\mathbf{B}$ direction is constrained by the electric field acceleration along $\mathbf{B}$ contrary to Aristotelian electrodynamics. \cite{gruzinov_aristotelian_2013} introduced two new quantities $E_0>0$ and $B_0$ according to the following invariants (but see also \cite{mestel_stellar_1999})
\begin{subequations}
	\begin{align}
	\mathcal{I}_1 & = E^2 - c^2 \, B^2 = E_0^2 - c^2 \, B_0^2 \\
	\mathcal{I}_2 & = \mathbf{E} \cdot \mathbf{B} = E_0 \, B_0 .
	\end{align}
\end{subequations}
Solving for the magnetic field strength~$B_0$ and keeping only the real solution with a positive sign~$+$ we get
\begin{equation}
B_0^2 = \frac{-\mathcal{I}_1 + \sqrt{\mathcal{I}_1^2+4\,c^2\,\mathcal{I}_2^2}}{2\,c^2} .
\end{equation}
In such a way, plugging this expression into the electromagnetic field transform given in eq.~(\ref{eq:transform_EB}) we find
\begin{subequations}
	\begin{align}
	\mathbf E' & = \frac{\Gamma \, E_0}{E_0^2/c^2 + B^2} \, \left[ \frac{E_0}{c^2} \, \mathbf E + B_0 \, \mathbf B \right] \\
	\mathbf B' & = \frac{\Gamma \, B_0}{E_0^2/c^2 + B^2} \, \left[ B_0 \, \mathbf B + \frac{E_0}{c^2} \, \mathbf E \right]
	\end{align}
\end{subequations}
They are therefore colinear because $E_0 \, \mathbf B' = B_0 \, \mathbf E'$. The frame velocity consequently simplifies into
\begin{equation}
\label{eq:FrameSpeed}
\mathbf V = \frac{\mathbf E \wedge \mathbf B}{E_0^2/c^2 + B^2} .
\end{equation}
Note in this expression the mixing between field strengths in both frames, one with subscript~$0$ and the other without any subscript. There exists however a symmetry in the sense that $E_0^2 + c^2 \, B^2 = E^2 + c^2 \, B_0^2$ so we can use either $E_0$ or $B_0$ but not both simultaneously. This velocity is always less than the speed of light if $\mathcal{I}_1\neq0$ and $\mathcal{I}_2\neq0$. Therefore, there always exist a frame where $\mathbf{E}$ and $\mathbf{B}$ are parallel, whatever the strength of $\mathbf{E}$ compared to $\mathbf{B}$. The vanishing magnetic field $\mathbf{B}=0$ or electric field $\mathbf{E}=0$  are special cases of the general treatment presented here the second one reducing to the electric drift motion.

The special case of a null field for which $\mathcal{I}_1 = \mathcal{I}_2 = 0$ is treated separately because then $E=c\,B$ and the speed of the frame is  exactly equal to~$c$ and thus is not a physical frame.

\subsection{Motion in the frame where $\mathbf{E}$ and $\mathbf{B}$ are parallel}

In the previous section, we showed that it is always possible to reduce the problem of particle motion into a configuration where $\mathbf{E}$ and $\mathbf{B}$ are parallel except for light like fields. In this frame, integration of the trajectory is particularly simple when expressed in terms of the proper time~$\tau$ of the particle. Any electromagnetic field configuration can always be reduced to a parallel electric and magnetic field by an appropriate change of reference. In order not to overload the notations, in this and the following subsections we omit the primes to designate the quantities expressed in the frame $K'$ where $\mathbf{E}$ and $\mathbf{B}$ are parallel. The integration of the equation of motion of a charged particle in such a field is relatively simple and straightforward. Indeed, let's consider an electromagnetic field such that $\mathbf{E}$ and $\mathbf{B}$ are directed along the $\mathbf{e}_z$ axis in a Cartesian coordinate system. The initial position of the particle is $(x_0, y_0, z_0)$ and its initial velocity is $\mathbf{v} = (v_x^0, v_y^0, v_z^0)$. The equation of motion in covariant form is
\begin{equation}
\frac{d p^i}{d\tau} = q \, F^{ik} \, u_k
\end{equation}
or in terms of the 4-velocity only
\begin{equation}
\frac{d u^i}{d\tau} = \frac{q}{m} \, F^{ik} \, u_k
\end{equation}
$u^i=(\gamma\,c,\gamma\,\mathbf{v})$ is the 4-velocity and $p^i=m\,u^i$ the 4-momentum.
In the Cartesian coordinate system, the electromagnetic field tensor is anti-diagonal and given by
\begin{equation}
F^{ik} =
\begin{pmatrix}
0 & 0 & 0 & -E/c \\
0 & 0 & -B & 0 \\
0 & B & 0 & 0 \\
E/c & 0 & 0 & 0
\end{pmatrix} .
\end{equation}
Introducing $\omega_E = \frac{q\,E}{m\,c}$ and $\omega_B = \frac{q\,B}{m}$, the equation of motion reduces to 
\begin{subequations}
	\begin{align}
	\frac{d u^0}{d\tau} & = \frac{q}{m} \, F^{03} \, u_3 = - \omega_E \, u_3 \\
	\frac{d u^1}{d\tau} & = \frac{q}{m} \, F^{12} \, u_2 = - \omega_B \, u_2 \\
	\frac{d u^2}{d\tau} & = \frac{q}{m} \, F^{21} \, u_1 = + \omega_B \, u_1 \\
	\frac{d u^3}{d\tau} & = \frac{q}{m} \, F^{30} \, u_0 = + \omega_E \, u_0 .
	\end{align}
\end{subequations}
All the components of the 4-velocity are brought back to their contravariant expressions so that by index elevation $u^i = \eta^{ik} \, u_k$. This implies that $u^0 = u_0$ for the temporal index and $u^a = - u_a$ for the spatial indices.
The system decouples into two size~2 subsystems each so that
\begin{subequations}
	\begin{align}
	\frac{d u^0}{d\tau} & = \omega_E \, u^3 \\
	\frac{d u^1}{d\tau} & = \omega_B \, u^2 \\
	\frac{d u^2}{d\tau} & = - \omega_B \, u^1 \\
	\frac{d u^3}{d\tau} & = \omega_E \, u^0 .
	\end{align}
\end{subequations}
Two variables are eliminated to reduce the system to the velocity components $u^0$ and $u^1$ only
\begin{subequations}
	\begin{align}
	\frac{d^2 u^0}{d\tau^2} & = \omega_E^2\, u^0 \\
	\frac{d^2 u^1}{d\tau^2} & = - \omega_B^2 \, u^1
	\end{align}
\end{subequations}
The general solutions are given by 
\begin{subequations}
	\begin{align}
	u^0 & = A \, e^{\omega_E\,\tau} + B \, e^{-\omega_E\,\tau} \\
	u^1 & = C \, \cos(\omega_B\,\tau) + D \, \sin(\omega_B\,\tau) .
	\end{align}
\end{subequations}
At initial time, we have $t = t_0$ corresponding to $\tau=0$ and $\mathbf v = v_0 \, \mathbf t$ thus $\mathbf u^0 = \gamma_0 \, (c , \mathbf v_0)$ with $\gamma_0=(1-(\beta_0)^2)^{-1/2}$. These initial conditions enforce
\begin{subequations}
	\begin{align}
	A & = \gamma_0 \, \frac{c+v_0^z}{2} \\
	B & = \gamma_0 \, \frac{c-v_0^z}{2} \\
	C & = \gamma_0 \, v_0^x \\
	D & = \gamma_0 \, v_0^y 
	\end{align}
\end{subequations}
thus the 4-velocity evolution given in terms of the proper time according to
\begin{subequations}
	\label{eq:VitesseParticule}
	\begin{align}
	u^0 & = \gamma_0 \, c \, \left[ \ch (\omega_E\,\tau) + \beta_0^z \, \sh  (\omega_E\,\tau) \right] \\
	u^3 & = \gamma_0 \, c \, \left[ \sh (\omega_E\,\tau) + \beta_0^z \, \ch  (\omega_E\,\tau) \right] \\
	u^1 & = \gamma_0 \, c \, \left[ \beta_0^x \, \cos (\omega_B\,\tau) + \beta_0^y \, \sin  (\omega_B\,\tau) \right] \\
	u^2 & = \gamma_0 \, c \, \left[ -\beta_0^x \, \sin (\omega_B\,\tau) + \beta_0^y \, \cos  (\omega_B\,\tau) \right] .
	\end{align}
\end{subequations}
All what remains is to integrate with respect to the proper time to find the trajectory of the particle
\begin{subequations}
	\label{eq:TrajectoireParticule}
	\begin{align}
	c \, ( t-t_0) & = \frac{\gamma_0 \, c }{\omega_E} \, \left[ \sh (\omega_E\,\tau) + \beta_0^z \, ( \ch (\omega_E\,\tau) - 1 ) \right] \\
	x - x_0 & = \frac{\gamma_0 \, c }{\omega_B} \, \left[ \beta_0^x \, \sin (\omega_B\,\tau) - \beta_0^y \, ( \cos (\omega_B\,\tau) - 1 ) \right] \\
	y - y_0 & = \frac{\gamma_0 \, c }{\omega_B} \, \left[ \beta_0^x \, ( \cos (\omega_B\,\tau) - 1 ) + \beta_0^y \, \sin (\omega_B\,\tau) \right] \\
	z - z_0 & = \frac{\gamma_0 \, c }{\omega_E} \, \left[ \ch (\omega_E\,\tau) - 1 + \beta_0^z \, \sh (\omega_E\,\tau) \right] .
	\end{align}
\end{subequations}
The trajectory is thus entirely determined as a function of proper time~$\tau$ in an analytical way with simple expressions including trigonometric and hyperbolic functions. These equations for 4-velocity and 4-position are implemented in the code.

%
Note that the equation of motion could be resolved immediately in the observer's frame of reference by diagonalizing the tensor of the electromagnetic field $F^{ik}$. This would be the same as the change of reference frame made above \citep{vandervoort_relativistic_1960}.

For testing the numerical algorithm and the limiting Lorentz factor due to numerical round off error, we check the implementation on a purely electric and a purely magnetic field.

\subsection{Vanishing magnetic field}

In the case of a vanishing magnetic field $\mathbf{B}=0$, the frame $K$ and $K'$ are identical and there is no need to boost from one frame to the other. The 4-velocity reduces to
\begin{subequations}
	\label{eq:no_magnetic_velocity}
	\begin{align}
	u^0 & = \gamma_0 \, c \, \left[ \ch (\omega_E\,\tau) + \beta_0^z \, \sh  (\omega_E\,\tau) \right] \\
	u^1 & = \gamma_0 \, c \, \beta_0^x \\
	u^2 & = \gamma_0 \, c \, \beta_0^y \\
	u^3 & = \gamma_0 \, c \, \left[ \sh (\omega_E\,\tau) + \beta_0^z \, \ch  (\omega_E\,\tau) \right]
	\end{align}
\end{subequations}
and the trajectory simply into
\begin{subequations}
	\label{eq:no_magnetic_position}
	\begin{align}
	c \, ( t-t_0) & = \frac{\gamma_0 \, c }{\omega_E} \, \left[ \sh (\omega_E\,\tau) + \beta_0^z \, ( \ch (\omega_E\,\tau) - 1 ) \right] \\
	x - x_0 & = \gamma_0 \, c \, \beta_0^x \, \tau \\
	y - y_0 & = \gamma_0 \, c \, \beta_0^y \, \tau \\
	z - z_0 & = \frac{\gamma_0 \, c }{\omega_E} \, \left[ \ch (\omega_E\,\tau) - 1 + \beta_0^z \, \sh (\omega_E\,\tau) \right] .
	\end{align}
\end{subequations}

\subsection{Vanishing electric field}

In the case of a vanishing electric field $\mathbf{E}=0$, the frame $K$ and $K'$ are again identical and there is no need to boost from one frame to the other. The 4-velocity reduces to
\begin{subequations}
	\label{eq:no_electric_velocity}
	\begin{align}
	u^0 & = \gamma_0 \, c \\
	u^1 & = \gamma_0 \, c \, \left[ \beta_0^x \, \cos (\omega_B\,\tau) + \beta_0^y \, \sin  (\omega_B\,\tau) \right] \\
	u^2 & = \gamma_0 \, c \, \left[ -\beta_0^x \, \sin (\omega_B\,\tau) + \beta_0^y \, \cos  (\omega_B\,\tau) \right] \\
	u^3 & = \gamma_0 \, c \, \beta_0^z
	\end{align}
\end{subequations}
All what remains is to integrate with respect to the proper time to find the trajectory of the particle
\begin{subequations}
	\label{eq:no_electric_position}
	\begin{align}
	( t-t_0) & = \gamma_0 \, \tau \\
	x - x_0 & = \frac{\gamma_0 \, c }{\omega_B} \, \left[ \beta_0^x \, \sin (\omega_B\,\tau) - \beta_0^y \, ( \cos (\omega_B\,\tau) - 1 ) \right] \\
	y - y_0 & = \frac{\gamma_0 \, c }{\omega_B} \, \left[ \beta_0^x \, ( \cos (\omega_B\,\tau) - 1 ) + \beta_0^y \, \sin (\omega_B\,\tau) \right] \\
	z - z_0 & = \gamma_0 \, c \, \beta_0^z \,\tau = v_0^z \, ( t-t_0) .
	\end{align}
\end{subequations}
Testing our algorithm in a constant and uniform electromagnetic field is meaningless because the solution is known analytically in any reference frame. Even the force-free field test introduced by \cite{ripperda_comprehensive_2018} 
is included in this exact analytical solution as demonstrated in the next paragraph. However, for numerical purposes, it is desirable to check the ability of our code to handle very high Lorentz factor above $\gamma=\numprint{e12}$ to look for round-off errors and possible issues related to finite digit precision.

\subsection{Motion in the frame where $\mathbf{E}$ and $\mathbf{B}$ are perpendicular}

When electric and magnetic fields are perpendicular, the second invariant vanishes $\mathcal{I}_2=0$, there exist always a frame where either the electric or the magnetic field vanishes depending on the sign of the invariant~$\mathcal{I}_1$. We are then back to the previous cases for a pure electric or magnetic field. Indeed, if $\mathcal{I}_1>0$ the constant $\alpha=c^2/E^2$ and therefore the velocity of the frame where $\mathbf{B}$ vanishes is
\begin{equation}
\mathbf{V}_B = \frac{\mathbf{E} \wedge \mathbf{B}}{E^2} \, c^2 .
\end{equation}
If $\mathcal{I}_1<0$ the constant $\alpha=1/B^2$ and therefore the velocity of the frame where $\mathbf{E}$ vanishes is
\begin{equation}
\mathbf{V}_E = \frac{\mathbf{E} \wedge \mathbf{B}}{B^2}
\end{equation}
that is the usual electric drift frame. Consequently, the cases $\mathcal{I}_1\neq0$ and $\mathcal{I}_2=0$ are included in the previous sections.

\subsection{Light like electromagnetic field}

Nevertheless, the case where both electromagnetic invariants vanish $\mathcal{I}_1 = \mathcal{I}_2 = 0$ must be treated separately because there exists no physical frame moving at a speed strictly less than~$c$ where either $\mathbf{E}$ or $\mathbf{B}$ vanishes. Going back to the equation of motion let us assume that the electric field is along $\mathbf{E} = E \, \ey$ and the magnetic field along $\mathbf{B} = B \, \ez$. The electromagnetic tensor then reduces to 
\begin{equation}
F^{ik} =
\begin{pmatrix}
0 & 0 & -E/c & 0 \\
0 & 0 & -B & 0 \\
E/c & B & 0 & 0 \\
0 & 0 & 0 & 0
\end{pmatrix} .
\end{equation}
The equation of motion therefore becomes
\begin{subequations}
	\begin{align}
	\frac{d u^0}{d\tau} & = \omega_E \, u^2 \\
	\frac{d u^1}{d\tau} & = \omega_B \, u^2 \\
	\frac{d u^2}{d\tau} & = \omega_E \, u^0 - \omega_B \, u^1 \\
	\frac{d u^3}{d\tau} & = 0 .
	\end{align}
\end{subequations}
The last equation integrates into $u^3=u^3_0 = constant$ and the trajectory is constraint to follow $z-z_0 = u^3_0\,(\tau-\tau_0)$.
Eliminating $u^0$ and $u^1$ from the equation evolving $u^2$ we find 
\begin{equation}
\frac{d^2u^2}{d\tau^2} + (\omega_B^2 - \omega_E^2) \, u^2 = 0 .
\end{equation}
Three cases must be distinguished
\begin{enumerate}
	\item a dominant electric field for $\omega_E > \omega_B$.
	\item a dominant magnetic field for $\omega_B > \omega_E$.
	\item a light like field if $\omega_E = \omega_B$.
\end{enumerate}
Note that $\omega_E^2 - \omega_B^2$ is related to the relativistic invariant $\mathcal{I}_1$ through $\frac{q^2}{m^2\,c^2} \, \mathcal{I}_1 = \omega_E^2 - \omega_B^2$.

In a light like electromagnetic field, there is no frame where $\mathbf{E}$ or $\mathbf{B}$ vanishes. A plane electromagnetic wave propagating in vacuum is a typical example. In this field, the two Lorentz invariants vanish $\mathcal{I}_1 = \mathcal{I}_2=0$. The equation for the 4-velocity reduces to
\begin{equation}
\frac{d^2 u^2}{d\tau^2} = 0
\end{equation}
whose solution is $u^2 = C \, \tau + \gamma_0\,v_0^y$ where $C$ is a constant of integration. The other two components are
\begin{subequations}
	\begin{align}
	u^0 & = \omega_E \, ( C \, \frac{\tau^2}{2} + \gamma_0\,v_0^y\,\tau) + \gamma_0\,c \\
	u^1 & = \omega_B \, ( C \, \frac{\tau^2}{2} + \gamma_0\,v_0^y\,\tau) + \gamma_0\,v_0^x .
	\end{align}
\end{subequations}
Replacing in $du^2/d\tau$ we find $C = \gamma_0 \, c \, (\omega_{\rm E} - \omega_{\rm B} \,\beta_0^x)$ because $\omega_E = \pm \omega_B$ and cannot be simplified further. The solution is therefore
\begin{subequations}
	\label{eq:light_like_velocity}
	\begin{align}
	u^0 & = \gamma_0 \, c \, [ 1 + \omega_{\rm E} \, (\omega_{\rm E} - \omega_{\rm B} \,\beta_0^x) \, \frac{\tau^2}{2} + \beta_0^y \, \omega_{\rm E} \, \tau ] \\
	u^1 & = \gamma_0 \, c \, [ \beta_0^x + \omega_{\rm B} \, (\omega_{\rm E} - \omega_{\rm B} \,\beta_0^x) \, \frac{\tau^2}{2} + \beta_0^y \, \omega_{\rm B} \, \tau ]  \\
	u^2 & = \gamma_0 \, c \, [ \beta_0^y + (\omega_{\rm E} - \omega_{\rm B} \,\beta_0^x) \, \tau ] \\
	u^3 & = \gamma_0 \, c \, \beta_0^z .
	\end{align}
\end{subequations}
This solution allows us to advance the particle velocity in time without taking care of the respective sign of $\omega_{\rm E}$ and $\omega_{\rm B}$.
Performing another integration gives the explicit form of the trajectory as
\begin{subequations}
	\label{eq:light_like_position}
	\begin{align}
	c \, (t-t_0) & = \gamma_0 \, c [ \tau + \omega_{\rm E} \, (\omega_{\rm E} - \omega_{\rm B} \,\beta_0^x) \, \frac{\tau^3}{6} + \beta_0^y \, \omega_{\rm E}\, \frac{\tau^2}{2} ] \\
	x - x_0 & = \gamma_0 \, c \, [ \beta_0^x \, \tau + \omega_{\rm E} \, (\omega_{\rm E} - \omega_{\rm B} \,\beta_0^x) \, \frac{\tau^3}{6} + \beta_0^y \, \omega_B\,\frac{\tau^2}{2}  ] \\
	y - y_0 & = \gamma_0 \, c \, [ \beta_0^y \, \tau  + (\omega_{\rm E} - \omega_{\rm B} \,\beta_0^x) \, \frac{\tau^2}{2} ] \\
	z - z_0 & = \gamma_0 \, c \, \beta_0^z \, \tau .
	\end{align}
\end{subequations}
These solutions must be included in the algorithm whenever both invariants vanish. Tests will be performed in a linearly and a circularly polarized plane electromagnetic wave in section~\ref{sec:Results}.

\subsection{Force-free field}

In the special case of a force-free field, the electric and magnetic field are perpendicular according to the expression
\begin{equation}
\label{eq:ForceFree}
\mathbf{E} + \mathbf{v} \wedge \mathbf{B} = \mathbf{0}
\end{equation}
where $\mathbf{v}$ is the particle velocity in the frame $K$. In that case, by construction $\mathbf{E} \cdot \mathbf{B} = 0$ and $E<c\,B$. The privileged frame~$K'$ according to the general transformation rule is therefore simply the electric drift frame moving at a velocity
\begin{equation}
\mathbf{V}_{\rm E} = \frac{\mathbf{E} \wedge \mathbf{B}}{B^2} = \mathbf{v} = \mathbf{V}_\parallel .
\end{equation}
Consequently the particle velocity~$\mathbf{v}$ is equal to the drift velocity~$\mathbf{V}_{\rm E}$ and equal to $\mathbf{V}_\parallel$ meaning that in the frame where $\mathbf{E}$ and $\mathbf{B}$ are parallel (actually $\mathbf{E}=0$) the particle is at rest. Thus no Lorentz force acts at all in $K'$ because initially $\mathbf{v}'=0$ and $\mathbf{E}'=0$, the particle remains at rest forever in $K'$ as long as the force-free condition eq.~(\ref{eq:ForceFree}) is satisfied, whatever the value of $\mathbf{B}'$. This result is analytically exact, there is no numerical error, even to round off accuracy, contrasting severely with other algorithms showing a slowly breaking of the force-free motion \citep{ripperda_comprehensive_2018}.

\subsection{Low electric and magnetic field}

The expression for particle trajectories given in eq.~(\ref{eq:TrajectoireParticule}) cannot be applied straightforwardly whenever $\omega_{\rm E}$ or $\omega_{\rm B}$ tends to zero because they appear in the denominator of some expressions. However, to include these cases in the general algorithm we designed, we performed a first order expansion whenever required to smoothly join the vanishing electric or magnetic field case presented previously.

In regions where the electric field is weak according to the condition $\omega_E \, \tau \ll 1$ the time and $z$ components are integrated according to
\begin{subequations}
	\label{eq:champ_faible_position}
	\begin{align}
	(t-t_0) & = \gamma_0 \, \tau \\
	z - z_0 & = v_0^z \, ( t-t_0) .
	\end{align}
	In regions where the magnetic field is weak according to the condition $\omega_B \, \tau \ll 1$ the $x$ and $y$ components are integrated according to
	\begin{align}
	x - x_0 & = \gamma_0 \, c \, \beta_0^x \, \tau \\
	y - y_0 & = \gamma_0 \, c \, \beta_0^y \, \tau .
	\end{align}
\end{subequations}
Taking into account these two limiting cases of vanishing electric or magnetic field, our algorithm can handle with any field strength and any geometry of the electromagnetic field as long as these fields are constant in time and uniform in space. Thus its robustness must be tested against spatially and temporally varying fields. This is checked in the next section. Nevertheless, our focus in this paper is about ultra-relativistic regime of particle motion, dealing with Lorentz factor as high as $\gamma = \numprint{e12}$. Although the analytical expressions are obviously valid for any $\gamma$, the finite precision of numerical implementation of any algorithm limits the possible range of Lorentz factor achievable. Therefore, we will also check our code on simple test cases such as purely electric or purely magnetic field in order to assess the stringent limitations of what can be done numerically with ultra-relativistic plasmas.

But before running simulations, we have to properly choose the optimal time step to evolve the particle motion. This crucial problem is addressed in the following section.

\section{Simulation time step determination}
\label{sec:TimeStep}

In the previous section, we showed that the equation of motion is most easily solved by introducing the particle proper time~$\tau$ and 4-velocity~$u$. Its time evolution could be performed by imposing and advancing the proper time. However, this is not the way numerical simulations are performed. We need to fix the observer time step~$\Delta t$ in the frame $K$ and not the proper time step~$\Delta \tau$ as measured by the particle in its rest frame.
In this section, we show that the proposed numerical scheme can be used in Particle-In-Cell simulations that is with a fixed observer time step~$\Delta t$. A PIC code uses discrete time steps in terms of the time in the frame of the simulation. The proposed scheme, on the other hand, uses discrete time steps in terms of the proper time~$\Delta \tau$ of each particle. Now, for a given time step in the frame of the simulation~$\Delta t$, we retrieve the corresponding time step in terms of proper time~$\Delta \tau$ for each particle. This requires to solve a non-linear equation. To make the scheme applicable to PIC simulations, we include a detailed discussion on this issue, deriving the equation to solve when calculating the time step in terms of proper time for each particle and how to solve it.

\subsection{Time step setting in different frames}

In a first step, the relation between particle proper time $\Delta \tau$ and the time~$\Delta t'$ in frame $K'$, eq.~(\ref{eq:TrajectoireParticule}) can be inverted analytically to give
\begin{equation}\label{eq:DeltaTau}
\omega_E \, \Delta \tau = \log \left( \frac{\xi + \sqrt{1-(\beta_0^z)^2 + \xi^2 }}{1+\beta_0^z} \right) = - \log \left( \frac{-\xi + \sqrt{1-(\beta_0^z)^2 + \xi^2 }}{1-\beta_0^z} \right) 
\end{equation}
where we introduced $\xi=\omega_E\,\Delta t'/\gamma_0 + \beta_0^z$. This expression is optimal for numerical computation, working also when $\beta_0^z$ is nearly +1 for the first expression and nearly -1 for the second expression. In a second step, the time $\Delta t'$ must be related to the observer time $\Delta t$ in frame $K$, the one used in the simulation to advance in time one time step. $\Delta t$ and $\Delta t'$ are related by a Lorentz boost similar to eq.~(\ref{eq:TransfoLorentz}) written most efficiently as
\begin{equation}\label{eq:DeltaT}
\Delta t = \Gamma \, \left( \Delta t' + \frac{\mathbf{V}\cdot \Delta \mathbf{r}'}{c^2} \right)
\end{equation}
where $\Gamma = (1-V^2/c^2)^{-1/2}$. 
The particle advance in position~$\Delta \mathbf{r}'$ in frame $K'$ is known analytically from eq.~(\ref{eq:TrajectoireParticule}) if $\Delta \tau$ is known. Symbolically, we write $\Delta \mathbf{r}'(\Delta \tau)$. Moreover, $\Delta \tau$ is found from $\Delta t'$ thanks to eq.~(\ref{eq:DeltaTau}) or symbolically $\Delta \tau(\Delta t')$ thus $\Delta t $ is found from solving the non-linear scalar equation~\ref{eq:DeltaT}, remembering that finally $\Delta \mathbf{r}'(\Delta t')$.

This procedure is best understood by a simple example. Let us consider a particle moving along an arbitrary direction in frame $K$ with constant speed~$\mathbf{v}$. In the frame $K'$, it moves at a constant speed $\mathbf{v}'$, $K'$ moving with respect to $K$ at a speed~$\mathbf{V}$. How to relate then the speeds $\mathbf{v}$ and $\mathbf{v}'$ knowing the trajectory in $K'$?

In this simple case, the vector position is given in frame $K'$ by $\Delta \mathbf{r}' = \mathbf{v}'\,\Delta t'$. Therefore $\Delta t'$ can be solved with respect to $\Delta t'$ as
\begin{equation}
\label{eq:DeltaTprime}
\Delta t' = \frac{\Delta t}{\Gamma \, (1 + \mathbf{V} \cdot \mathbf{v}' / c^2)}
\end{equation}
$\Delta \mathbf{r}$ is then related to $\Delta \mathbf{r}'$ via the spatial part of the Lorentz transformation. In such a way, the velocity 
\begin{equation}
\label{eq:Vitesse}
\mathbf{v} = \frac{\Delta \mathbf{r}}{\Delta t} = \frac{\mathbf{v}'}{\Gamma \, (1 + \mathbf{V} \cdot \mathbf{v}' / c^2)} + \left( 1 + \frac{\Gamma}{\Gamma + 1} \, ( \mathbf{V} \cdot \mathbf{v}' / c^2 ) \right) \,  \frac{\mathbf{V}}{1 + \mathbf{V} \cdot \mathbf{v}' / c^2}
\end{equation}
is computed from the velocity in the frame $K'$. It can be checked that the results agree with the relativistic composition of velocities. The crucial point in this derivation is the analytical inversion of the relation $\Delta t(\Delta t')$ into $\Delta t'(\Delta t)$. This is no more the case for a particle in an electromagnetic field. We therefore have to resort to numerical inversion by some root finding methods like Newton-Raphson scheme or other techniques as described for instance in \cite{press_numerical_2007}.

\subsection{Adaptive time steps}

The optimal time step strongly depends on the local value of the electromagnetic field and on the particle velocity. Due to the Lorentz transformation of the electromagnetic tensor and because of time dilation, for ultra-relativistic particles, we expect a significant gain in time computation when this time step is adaptively adjusted to the local fields and particle velocities. In order to improve and better control the accuracy of our numerical solution from time step to time step, we implemented also a semi-implicit iterative scheme where the actual constant electromagnetic field employed to advance the particle is the one located midway between the time and position of the particle at the instant $t^n$ and $t^{n+1}$. It resembles the algorithm we already used in \cite{petri_fully_2017}. If the solution does not converge to a prescribed precision after a limited number of iteration~$N_{\rm max}$ (we set it to $N_{\rm max}=10$), the time step is decreased by a factor~2 and the process starts again with at most $N_{\rm max}$~iterations. If necessary the time step is again diminished by a factor~2 until convergence is reached. Figure~\ref{fig:code2} summarized the pseudo-code used to advance the particle position one time step by using the adaptive scheme.

\tikzstyle{decision} = [diamond, draw, fill=blue!20, 
text width=4.5em, text badly centered, node distance=2cm, inner sep=0pt]
\tikzstyle{block} = [rectangle, draw, fill=blue!20, 
text width=6em, text centered, rounded corners, minimum height=4em]
\tikzstyle{line} = [draw, -latex']
\tikzstyle{cloud} = [draw, ellipse,fill=red!20, node distance=2cm,
minimum height=2em]

\begin{figure}
\centering
\begin{tikzpicture}[node distance = 2cm, auto]
\node [block] (init) {$\mathbf{x}^n,\mathbf{u}^n$\\$N_{\rm max}, \epsilon$};
\node [block, right of=init, node distance=3cm] (solutions) {$\mathbf{F}^*=\mathbf{F}^n$\\$i=0$};
\node [block, below of=solutions] (identify) {analytical solutions\\ with $\mathbf{F}^*, \Delta\tau$};
\node [block, below of=identify] (evaluate) {guess $\mathbf{x}^{n+1},\mathbf{u}^{n+1}$\\$\mathbf{F}^{n+1}$\\it++};
\node [block, below of=evaluate] (midway) {$\mathbf{F}^*=\mathbf{F}\left(\frac{\mathbf{x}^{n}+\mathbf{x}^{n+1}}{2}\right)$};
\node [block, left of=evaluate, node distance=3cm] (update) {$i=0$\\$\Delta \tau = \Delta \tau /2$};
\node [decision, below of=midway] (decide) {$i>N_{\rm max}$?};
\node [decision, right of=evaluate, node distance=3cm] (boucle) {$\|\Delta\mathbf{F}\|>\epsilon$};
\node [block, right of=boucle, node distance=3cm] (stop) {$\mathbf{x}^{n+1},\mathbf{u}^{n+1}$};
\node [block, right of=decide, node distance=3cm] (erreur) {$\Delta \mathbf{F} = \mathbf{F}_i^{n+1} - \mathbf{F}_{i-1}^{n+1}$};
\path [line] (init) -- (solutions);
\path [line] (solutions) -- (identify);
\path [line] (identify) -- (evaluate);
\path [line] (evaluate) -- (midway);
\path [line] (midway) -- (decide);
\path [line] (decide) -| node [near start] {yes} (update);
\path [line] (update) |- (identify);
\path [line] (decide) -- node [near start] {no}(erreur);
\path [line] (boucle) |- node [near start] {yes}(identify);
\path [line] (erreur) -- (boucle);
\path [line] (boucle) -- node {no}(stop);
\end{tikzpicture}
\caption{\label{fig:code2}Pseudo-code summarizing the Picard iteration scheme to advance on time step. $\mathbf{F}$ means that it must apply to both fields $\mathbf{E}$ and $\mathbf{B}$.}
\end{figure}
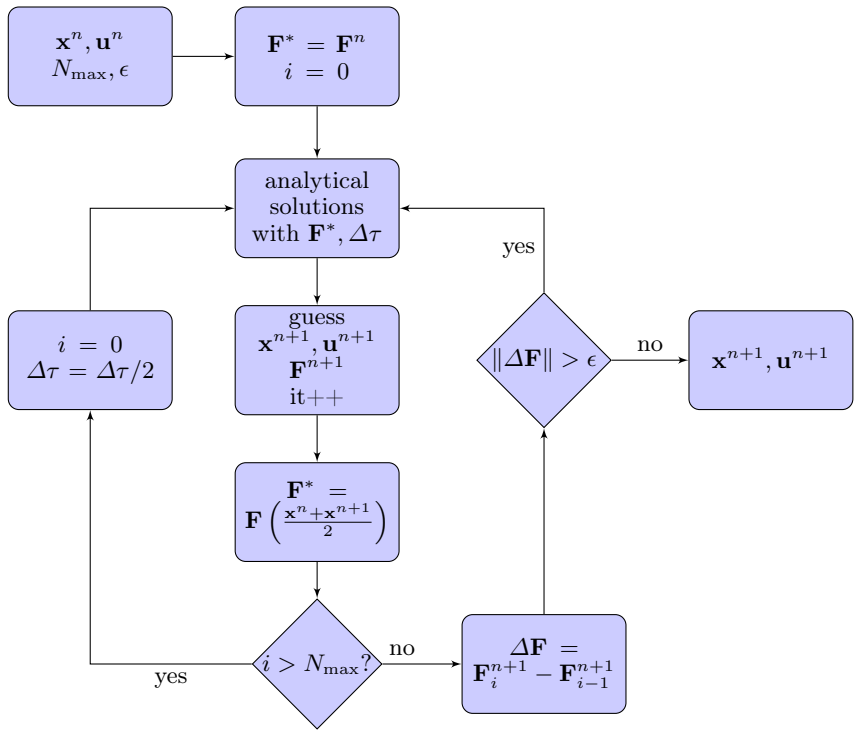

Having an algorithm to impose a fixed or prescribed \textit{observer time step}, we can couple the particle orbit integrator to the field advance like any other existing PIC code. There is no special care to first advance particles and second to compute the new fields. Let us now look at some particular tests of our algorithm.

\section{Numerical tests in spatially and temporally varying fields}
\label{sec:Results}

We extensively tested our new algorithm first against trivial configurations of an uniform electric or magnetic field, then in a cross electric and magnetic field following the electric drift frame. More stringent tests like the electrostatic Kepler problem, the magnetic gradient drift and motion in an ultra-strong linearly or circularly polarized plane wave are also considered.

\subsection{Normalisation and adimensionalisation}

Before showing some numerical results, we normalize the relevant quantities to characteristic values. As we deal with relativistic motion, the speed is conveniently normalized to the speed of light~$c$. Next we introduce a characteristic frequency~$\omega$ related the electromagnetic frequencies $\omega_{\rm B}$, $\omega_{\rm E}$ or the plasma frequency~$\omega_{\rm p}$ depending on the problem studied. This naturally leads to a characteristic length given by $c/\omega$. Moreover, a normalized time is then introduced by $\tilde{t} = \omega\,t$. In the simulation results shown below, we will always refer to these normalised quantities and plot graphs according to this convention. Concretely, for simulation purposes, we specify the charge~$q$ and mass~$m$ of the particle in the different tests such that $q=1$ and $m=1$ if not otherwise specified.

\subsection{Purely electric field}

Let us start with a homogeneous and uniform electric field. Take as an initial condition $t_0=0$ and the position of the particle to be $x_0 = y_0 = z_0 = 0$ with no initial velocity such that $\boldsymbol{\beta} = \mathbf{0}$ and thus $\gamma_0=1$. The particle world line is therefore a straight line parametrized with respect to its proper time~$\tau$ according to
\begin{subequations}
	\begin{align}
	\omega_E \, t & = \sh (\omega_E\,\tau) \\
	x & = 0 \\
	y & = 0 \\
	\omega_E \, z & = c \, \left[ \ch (\omega_E\,\tau) - 1 \right] .
	\end{align}
\end{subequations}
Expressed in terms of the observer time~$t$, the trajectory and velocity become
\begin{subequations}
	\begin{align}
	\omega_E \, z & = c \, \left[ \sqrt{ 1 + (\omega_E \, t)^2} - 1 \right] \\
	\frac{v_z}{c} & = \frac{\omega_E \, t}{\sqrt{ 1 + (\omega_E \, t)^2}} .
	\end{align}
\end{subequations}
The Lorentz factor grows with time~$t$ according to
\begin{equation}
\gamma = \ch (\omega_E\,\tau) = \sqrt{ 1 + (\omega_E \, t)^2} .
\end{equation}
Note that the only relevant time scale in this problem is the normalized quantity $\tilde{t} = \omega_E \, t$ or $\tilde{\tau} = \omega_E\,\tau$.

\begin{figure}
	\centering
	\includegraphics{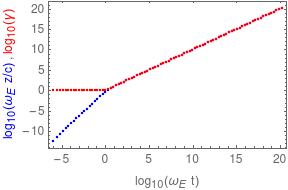}
	\caption{Position~$z$ in blue and Lorentz factor in red of an electron in the accelerating electric field $\mathbf{E} = E \, \mathbf{e}_{\rm z}$ and with increasing observer time step~$\Delta t$.}
	\label{fig:position_electrique_pur}
\end{figure}

An example of accelerating electric field is shown in fig.~\ref{fig:position_electrique_pur} for the position in blue and for the Lorentz factor in red. What matters is not the particle charge and mass, but the quantity $q\,E/m\,c = \omega_{\rm E}$ which gives the typical time scale for acceleration. The time step in the observer frame~$\Delta t$ is imposed by the user. We took an initial value of $\omega_E \, \Delta t = 10^{-6}$ although there is no restriction on $\Delta t$. Indeed, after each iteration we multiplied it by 2 in order to show the flexibility of adapting the \textit{observer time step}. Then the proper time step $\Delta\tau$ is computed according to eq.~(\ref{eq:DeltaTau}). The particle first accelerates in the Newtonian regime with a quadratic increase in position~$z$ up to the point where it reaches almost the speed of light. After a time $\omega_E \, t\gtrsim1$ it goes at almost constant speed $v_{\rm z}\approx c$. The Lorentz factor then increases almost linearly with time $\gamma \approx \omega_E \, t$. We let the particle gain energy up to $\gamma = \numprint{e20}$ to check possible issues related to numerical round off and truncation. No special problems were met for these ultra-relativistic speeds.

\subsection{Purely magnetic field}

\begin{figure}
	\centering
	\includegraphics{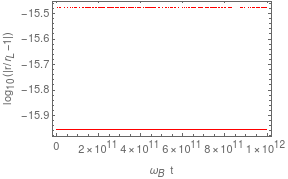}
	\caption{Evolution of the error in the orbital radius of an electron in the uniform magnetic field with $\gamma = \numprint{e10}$.}
	\label{fig:rayon_magnetique_pur}
\end{figure}

Let us go on with a homogeneous and uniform magnetic field. Take as an initial condition $t_0=0$ and the position of the particle in the $xOy$ plane such that $x_0=z_0=0$ and $y_0=\gamma_0 \, v_0^x/\omega_B = r_{\rm L}$ with initial velocity such that $\boldsymbol{\beta} = (\beta_0^x,0,\beta_0^z)$. $r_{\rm L} = \gamma\,v/\omega_{\rm B}$ is the Larmor radius of the trajectory. The Lorentz factor is constant and given by $\gamma=\gamma_0$. The particle world line is therefore
\begin{subequations}
	\begin{align}
	t & = \gamma_0 \, \tau \\
	x & = r_{\rm L} \, \sin (\omega_B\,\tau) \\
	y & = r_{\rm L} \, \cos (\omega_B\,\tau) \\
	z & = v_0^z \, t .
	\end{align}
\end{subequations}
The particle gyro-frequency is $\omega_B$ in proper time but reduced to $\omega_B/\gamma_0$ in the observer frame as is well known from special relativity. In the observer frame, the relevant normalized time scale is therefore $\omega_{\rm B}\,t$. For an ultra-relativistic particle with $v_0^x \approx c$, its Larmor radius is $r_{\rm L} \approx \gamma_0 \, c/\omega_B$, the expression used in the introduction. An example is shown in Fig.~\ref{fig:rayon_magnetique_pur} for $B=1$ and $\gamma = \numprint{e10}$. The time step is set to $\omega_B \, \Delta \tau = \numprint{e-2}$. The radius of the orbit stay at $r_{\rm L}$ to very high accuracy, more than 15~digits of precision. The Lorentz factor remains constant and equal to $\gamma_0$ as expected.

\subsection{Cross electric and magnetic fields}

\begin{figure}
	\centering
	\includegraphics{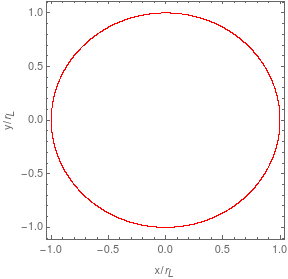}
	\caption{Gyromotion of an electron in the electric drift frame with  $\Gamma_{\rm E} = \numprint{e3}$ and $\gamma=\numprint{e10}$. The Larmor radius is $r_{\rm L}=\numprint{e10}$.}
	\label{fig:cercle_drift_pur}
\end{figure}

\begin{figure}
	\centering
	\includegraphics{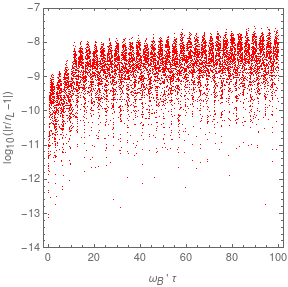}
	\caption{Relative error in the Larmor radius computed from the transformation in eq.(\ref{eq:DriftTransform}). The proper time~$\tau$ is shown on the $x$-axis normalized to the cyclotron frequency in the electric drift frame~$\omega_{\rm B'}$. }
	\label{fig:rayon_drift_pur}
\end{figure}

As another proof of the efficiency of the algorithm, we compute the trajectories in a crossed electromagnetic field ($\mathbf{E} \cdot \mathbf{B} = 0$) where the average motion is an electric drift in the $\mathbf{E} \wedge \mathbf{B}$ direction at the electric drift speed $\mathbf{v}_{\rm E} = \mathbf{E} \wedge \mathbf{B}/B^2$. In the frame moving at $\mathbf{v}_{\rm E}$ the electric field vanishes and the particle simply follows an helicoidal motion along the constant magnetic field, see the case treated in the previous paragraph. This motion is only allowed for weak electric fields satisfying $E<c\,B$ thus enforcing $v_{\rm E}<c$. For concreteness, let us assume an electric field directed along $\ey$ and a magnetic field directed along $\ez$. The electric drift speed becomes $\mathbf{v}_{\rm E} = (E/B) \, \ex$. A Lorentz transformation of the electromagnetic field with Lorentz factor $\Gamma_{\rm E} = 1/\sqrt{1-v_{\rm E}^2/c^2}$ along $\mathbf{v}_{\rm E}$ shows that in the comoving frame
\begin{subequations}
	\begin{align}
	\mathbf{E}' & = 0 \\
	\mathbf{B}' & = \mathbf{B}/\Gamma_{\rm E} .
	\end{align}
\end{subequations}
The electric field vanishes as expected and the magnetic field is decreased potentially by a large ratio equal to the Lorentz factor of the comoving frame. As initial conditions for the particle position and velocity, we choose a helicoidal motion as explained in the previous paragraph and corresponding to an evolution in the solely magnetic field $\mathbf{B}'$ as seen in the electric drift frame. These quantities are then transformed to the observer inertial frame according to Lorentz transformations for the velocity $\mathbf{v}$ of the particle. The algorithm is checked by computing the particle trajectory in the drift frame and the corresponding Lorentz factor of the particle that should remain constant in that frame. Using the Lorentz transformation the coordinates in the drift frame are
\begin{subequations}
	\label{eq:DriftTransform}
	\begin{align}
	x' & = \Gamma_{\rm E} \, ( x - v_{\rm E} \, t ) \\
	y' & = y \\
	z' & = z .
	\end{align}
\end{subequations}
For numerical purposes, the intensity of the electric field is set such that $\Gamma_{\rm E}=\numprint{e3}$, the particle Lorentz factor in this drift frame is $\gamma=\numprint{e10}$ and that of the magnetic field is $B=1$. Typical results are depicted in fig.~\ref{fig:cercle_drift_pur} for the trajectory in the electric drift frame which is usually an helicoidal motion and here exactly a circle in the comoving plane $x'O'y'$. 
The trajectory projected onto the $x'O'y'$ plane remains a circle to very good accuracy with no change in radius within 8~digits, see Fig.~\ref{fig:rayon_drift_pur}. The accuracy seems less good than in the previous case for a purely magnetic field. We lost several digits in the computation of the Larmor radius. This loss of precision is imputed to the procedure used to evaluate the trajectory in the drift frame. Indeed, a Lorentz transform is required for the particle position as explained in eq.(\ref{eq:DriftTransform}). This coordinate transform induced an additional error to the Larmor radius estimate. We lose several digits during the subtraction for $x'$. We checked that the error does not decrease with decreasing time step.

The most general and realistic fields are spatially and temporally varying. In these cases, the electromagnetic field has to be determined at some position and time during the particle motion. As severe tests of our algorithm, we study four problems of which three have exact analytical relativistic solutions for the particle trajectory. The relativistic electrostatic Kepler two body problem of an electron orbiting around a fixed positive ion is an interesting test for a spatially varying electric field. It is a central force case. The other two solutions correspond to a particle moving in a plane electromagnetic wave linearly or circularly polarized, a light like electromagnetic field. A last less trivial example is depicted by a particle drifting in the equatorial plane of a static magnetic dipole. We discuss in depth these regimes in the following paragraphs. The analytical solutions are described in \cite{uzan_theories_2014} and \cite{gourgoulhon_relativite_2010}. For completeness we recall them in the following paragraphs.

\subsection{Central electric force}

\begin{figure}
	\centering
	\includegraphics{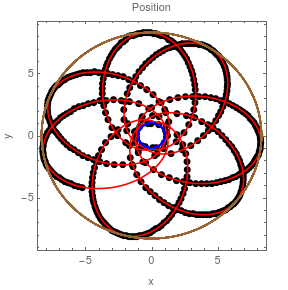}
	\caption{Motion of an electron in the electric field of a fixed proton, black points. The relativistic precession of the orbit is clearly visible. The minimal and maximum radius of the orbit as predicted by the analytical formulae are shown by two circles tangent to the trajectory. A piece of the exact analytical solution is also shown in red.}
	\label{fig:position_kepler_electrique}
\end{figure}

\begin{figure}
	\centering
	\includegraphics{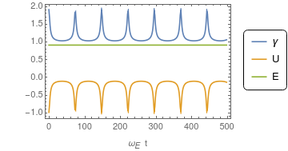}
	\caption{Total energy~$\mathcal{E}$, relativistic kinetic energy~$\gamma\,m\,c^2$ and electrostatic potential energy~$U$ of an electron in the electric field of a fixed proton. $\mathcal{E}$ is accurately conserved, being a constant of motion.}
	\label{fig:lorentz_kepler_electrique}
\end{figure}

\begin{figure}
	\centering
	\includegraphics{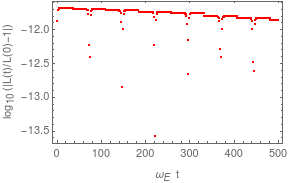}
	\caption{Relative error in the angular momentum~$L$. $L$ is conserved within 11~digits.}
	\label{fig:moment_kepler_electrique}
\end{figure}

\begin{figure}
	\centering
	\includegraphics{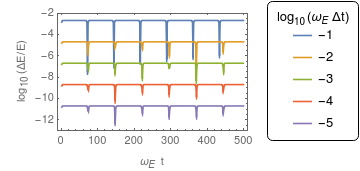}
	\caption{Relative error in the total energy depending on time steps~$\Delta t$.}
	\label{fig:erreur_kepler_electrique}
\end{figure}

The two body problem in gravitational physics can be transposed in an equivalent electrostatic problem including relativistic velocity. In this case, a particle with charge $q$ and mass $m$ orbits around a fixed central particle with charge~$Q$. For bounded orbits we require $q\,Q<0$. Solutions are given by conservation of energy~$\mathcal{E}$ (not to be confused with the electric field strength in this particular example) and angular momentum~$L$. The electric force applied to the orbiting particle is 
\begin{equation}
\mathbf{f} = \frac{q\,Q}{4\,\pi\,\varepsilon_0\,r^3} \, \mathbf{r} .
\end{equation}
The orbital motion stays in a plane that we choose as the $xOy$ plane. 
For $\left|q\,Q\right|<4\,\pi\,\varepsilon_0\,L\,c$, the general solution is 
\begin{subequations}
	\begin{align}
	r(t) & = \frac{p}{1 + e \, \cos(\Omega_{\rm p} \, (\varphi(t) - \omega))} \\
	\Omega_{\rm p} & = \sqrt{1 - \left( \frac{q\,Q}{4\,\pi\,\varepsilon_0\,L\,c}\right)^2 } \\
	p & = \frac{\Omega_{\rm p}^2}{- \frac{q\,Q}{4\,\pi\,\varepsilon_0\,c^2} \, \frac{\mathcal{E}}{L^2}} \\
	e^2 & = \frac{1}{\mathcal{E}^2} \, \left[ m^2\,c^4 + \frac{\mathcal{E}^2 - m^2\,c^4}{\left( \frac{q\,Q}{4\,\pi\,\varepsilon_0\,L\,c}\right)^2} \right] .
	\end{align}
\end{subequations}
The particle trajectory is plane and therefore described in a cylindrical coordinate system~$(r,\varphi)$ by the parametric function $r(\varphi)$. It depends implicitly on time~$t$ because of $\varphi$ being a function of time~$t$ (the solution for $\varphi$ is not shown here). Note also that $\Omega_{\rm p}<1$ meaning that the path is not a closed curved but a prograde precessing ellipse. $p$ is called the orbital parameter and is related to the semi-major axis~$a$ and eccentricity of the ellipse via $p=a\,(1-e^2)$. The periastron is located at a phase~$\omega$ with respect to the $x$-axis. These notations are common when studying stellar orbits in binary systems. This initial phase~$\omega$ is deduced from the initial condition~$r=r_0$ at $\varphi=\varphi_0$. Explicitly, we find
\begin{equation}
\Omega_{\rm p} \, (\varphi_0 - \omega) = \arccos \left( \frac{p/r_0 - 1}{e} \right) .
\end{equation}
An example of relativistic particle trajectory showing the precession of the orbit is given in fig.~\ref{fig:position_kepler_electrique}. A piece of the exact analytical solution is also shown and matches perfectly the output of the numerical simulations. The total energy~$\mathcal{E}$ is split into relativistic kinetic energy~$\gamma\,m\,c^2$ and electrostatic potential energy~$U$. Inspection of fig.~\ref{fig:lorentz_kepler_electrique} demonstrates that the total energy is accurately conserved during time evolution. Angular momentum is also conserved to high accuracy, at least 6~digits, see the relative error evolving in time in fig.\ref{fig:moment_kepler_electrique}.
Finally, the relative error in the total energy $\Delta \mathcal{E}/\mathcal{E}$ depending on time step is shown in fig.~\ref{fig:erreur_kepler_electrique}. We conclude that the scheme is second order in time, decreasing the error according to $\Delta \mathcal{E}/\mathcal{E} \propto \Delta t^{2}$.

\subsection{Magnetic drift in a dipole}

The magnetic field created by a dipole of magnetic dipole moment~$\boldsymbol{\mu}$ is given by the expression
\begin{equation}\label{eq:dipole_magnetique}
\mathbf{B} = \frac{\mu_0}{4\,\pi\,r^3} \, \left( \frac{3 \, (\boldsymbol{\mu}\cdot\mathbf{r}) \, \mathbf{r}}{r^2} - \boldsymbol{\mu} \right)
\end{equation}
where $\mu_0$ is the magnetic permeability. 
For a magnetic moment aligned with the $z$-axis, the magnetic field in the equatorial plane $xOy$ is purely vertical and given by
\begin{equation}\label{eq:dipole_equateur}
B_{\rm z} = B\, \frac{R^3}{r^3}
\end{equation}
where $B$ is the field strength at a distance $R$ from the origin.

For non-relativistic particles, the magnetic gradient drift velocity depending on the perpendicular velocity~$v_\perp$ is given by \citep{baumjohann_basic_1996}
\begin{equation}\label{eq:vitesse_derive}
\mathbf{v}_{\nabla\rm B} = \frac{m\,v_\perp^2}{2\,q\,B} \, \frac{\mathbf{B} \wedge \nabla B}{B^2} .
\end{equation}
For the above magnetic dipole geometry, in the equatorial plane we find a magnetic gradient velocity of
\begin{equation}\label{eq:VD}
\mathbf{v}_{\nabla\rm B} = \mp \frac{3}{2} \, \frac{m\,v_\perp^2}{q\,B\,r} \, \ephi = \mp \frac{3}{2} \, \frac{v_\perp^2}{\omega_{\rm B}\,r} \, \ephi =  \mp \frac{3}{2} \,v_\perp \, \frac{r_{\rm L}}{r} \, \ephi
\end{equation}
solely directed into the azimuthal direction and decreasing with distance as~$1/r$.

\begin{figure}
	\centering
	\includegraphics{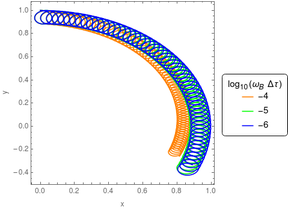}
	\caption{Orbit of an electron in the equatorial plane of a magnetic dipole for $B=\numprint{e3}$ and $\gamma \approx \numprint{71}$.}
	\label{fig:position_drift_pur_a1}
\end{figure}

\begin{figure}
	\centering
	\includegraphics{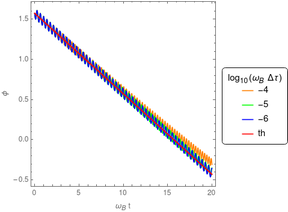}
	\caption{Observed and expected drift speed of a positron in the equatorial plane of a magnetic dipole for $B=\numprint{e3}$ and $\gamma \approx \numprint{71}$.}
	\label{fig:position_drift_pur_a1_vgb}
\end{figure}

Magnetic gradient drift motion is important in fields significantly varying on a length scale comparable to the Larmor radius. We study such motion in the equatorial plane of a magnetic dipole. We known that the particle must gyrate around the origin where the dipole is located. This motion is induced by the magnetic gradient drift in the azimuthal direction. The particle stays within two circles. An example of this drift motion is shown in fig.~\ref{fig:position_drift_pur_a1} for $B=\numprint{e3}$ and $\gamma \approx \numprint{71}$. The time step is varied and taken such that $\log_{10} (\omega_{\rm B} \, \Delta \tau) = \{0,-1,-2,-3\}$. The Lorentz factor is conserved as is easily checked. 
The observed drift speed is compared to the expected drift speed in fig.~\ref{fig:position_drift_pur_a1_vgb}. For a gyromotion well resolved in the proper time ($\omega_{\rm B} \, \Delta \tau\ll1$), the expected trajectory and drift speed are well reproduced.

\begin{figure}
\centering
\includegraphics{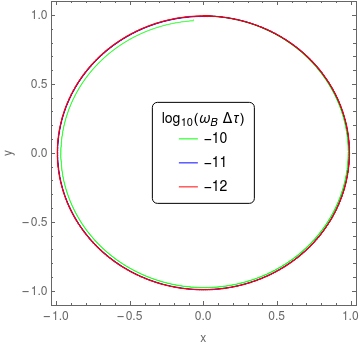}
\caption{Orbit of the guiding centre of an electron in the equatorial plane of a magnetic dipole for $B=\numprint{e8}$ and $\gamma \approx \numprint{e6}$.}
\label{fig:position_drift_pur_a3}
\end{figure}

\begin{figure}
\centering
\includegraphics{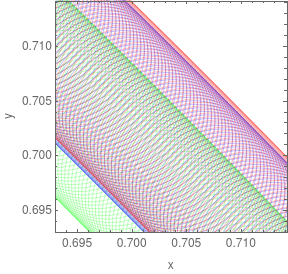}
\caption{Zoom into the orbit of an electron in the equatorial plane of a magnetic dipole for $B=\numprint{e8}$ and $\gamma \approx \numprint{e6}$.}
\label{fig:position_drift_pur_a3_zoom}
\end{figure}

\begin{figure}
	\centering
	\includegraphics{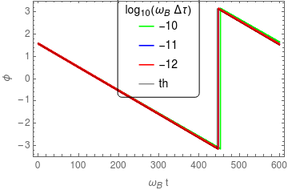}
	\caption{Observed and expected drift speed of a positron in the equatorial plane of a magnetic dipole for $B=\numprint{e8}$ and $\gamma \approx \numprint{e6}$.}
	\label{fig:position_drift_pur_a3_vgb}
\end{figure}

A second more stringent example is given by $B=\numprint{e8}$ and $\gamma \approx \numprint{e6}$. The corresponding trajectory is shown in fig.~\ref{fig:position_drift_pur_a3} for its guiding centre. The particle gyrates at very high frequency, turning millions of rotations before significantly translating its guiding centre a distance of the order its Larmor radius. To get the true trajectory without stroboscopic effects, we reduce the time step to $\Delta \tau=\numprint{e-9}$. A zoom into the plot show clearly the gyration, fig~\ref{fig:position_drift_pur_a3_zoom}. However, in such a case, it is absolutely useless the resolve the gyration frequency. The time step can be increased by several orders of magnitude without losing accuracy. This is the main advantage of our new particle solver, a key feature to solve kinetic problems around strongly magnetized neutron stars.
The drifting velocity also corresponds to the expected value, Fig.~\ref{fig:position_drift_pur_a3_vgb}.

We finish our extensive test of the algorithm by considering two time varying electromagnetic field configurations represented by linearly and circularly polarized plane waves.

\subsection{Linearly polarized plane wave}

\begin{figure}
	\centering
	\includegraphics{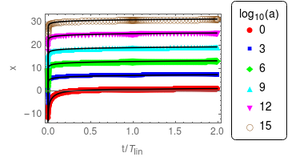}
	\caption{Motion of an electron in a linearly polarized plane wave for different strength parameters $a=10^i$ with $i\in\{0,3,6,9,12,15\}$. Symbols correspond to the analytical solution given in eq.~(\ref{eq:position_onde_lineaire}).}
	\label{fig:onde_lineaire}
\end{figure}

\begin{figure}
	\centering
	\includegraphics{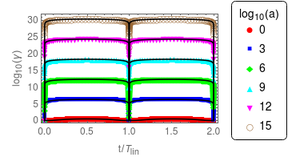}
	\caption{Lorentz factor of an electron in a linearly polarized plane wave for different strength parameters $a=10^i$ with $i\in\{0,3,6,9,12,15\}$. Symbols correspond to the analytical solution eq.~(\ref{eq:vitesse_onde_lineaire}).}
	\label{fig:onde_lineaire_gamma}
\end{figure}

Consider a linearly polarized plane wave propagating along the $\ex$ direction such that the 4-vector potential is $A^\alpha=(0,0,\frac{E}{\omega}\,\cos\xi,0)$. The wave vector is therefore $K^\alpha=(\frac{\omega}{c},k,0,0)$ from which we deduce the phase $\xi=\omega\,t-k\,x$. The electromagnetic field is then given by 
\begin{subequations}
	\begin{align}
	\mathbf{E} & = E \, \sin \xi \, \ey \\
	\mathbf{B} & = \frac{E}{c} \, \sin \xi \, \ez .
	\end{align}
\end{subequations}
Initially the particle is at rest with a 4-velocity $u^\alpha_0=(c,\mathbf 0)$. Introducing the strength parameter of the wave by the ratio
\begin{equation}
a = \frac{q\,E}{m\,c\,\omega}
\end{equation}
the 4-velocity has components
\begin{subequations}
	\label{eq:vitesse_onde_lineaire}
	\begin{align}
	u^x & = \frac{a^2}{2} \, c \, (\cos \xi - 1 )^2 \\
	u^y & = -a \, c \, (\cos \xi - 1 ) \\
	\label{eq:vitesse_onde_lineaire_0}
	u^0 & = c + u^x .
	\end{align}
\end{subequations}
More generally, exact analytical solutions for plane electromagnetic waves in vacuum (linearly or circularly polarized) have been derived. The methodology using 4-vectors and tensors can be found for instance in \cite{uzan_theories_2014}. See also \cite{michel_electrodynamics_1999} for typical applications to pulsars. For brevity, we do not reproduce these computations in this work.
The mean spatial velocity becomes 
\begin{subequations}
	\begin{align}
	<v^x> & = \frac{3}{4} \, \frac{a^2\,c}{1+3\,a^2/4} \\
	<v^y> & = \frac{a\,c}{1+3\,a^2/4} .
	\end{align}
\end{subequations}
After integration, assuming the particle starts at rest at the origin at $\xi=0$, we find
\begin{subequations}
	\label{eq:position_onde_lineaire}
	\begin{align}
	\omega \, x & = \frac{a^2\,c}{8} \, (6\,\xi - 8 \, \sin\xi + \sin 2\,\xi) \\
	\omega \, y & = a\,c\, (\xi - \sin\xi) \\
	\omega \, c \, t & = c\,\xi + \omega \, x .
	\end{align}
\end{subequations}
The particle motion is completely described by the strength parameter~$a$ (disregarding the initial conditions that are not part of the physical parameters). Examples of motion along the~$x$ axis are shown in fig.~\ref{fig:onde_lineaire} for a series of mildly and ultra-relativistic strength parameters $a=10^i$ with $i\in\{0,3,6,9,12,15\}$. The mean motion with average velocity $<v_x>$ is also shown as black solid lines. The associated Lorentz factor time evolution is shown in fig.~\ref{fig:onde_lineaire_gamma}. The numerical integration is compared to the analytical solution depicted by coloured symbols. Both are in perfect agreement. We are able to simulate acceleration to Lorentz factors well above $\gamma = \numprint{e12}$. This is compulsory to faithfully study lepton acceleration in neutron star magnetospheres.
The true motion in this plane wave is perfectly periodic with a period given by
\begin{equation}\label{eq:periode_lineaire}
T_{\rm lin} = 2\,\pi\,\left( 1 + \frac{3}{4} \, a^2 \right)
\end{equation}
the particle returning to rest at each period~$T_{\rm lin}$. The maximum Lorentz factor, reached at $t=T_{\rm lin}/2$ is
\begin{equation}\label{eq:gamma_max}
\gamma_{\rm max} = 1 + 2 \, a^2.
\end{equation}
\cite{arefiev_temporal_2015} performed similar simulations in a strong plane electromagnetic wave and found significant discrepancies between analytical and numerical results already with strength parameters $a\approx25$. Moreover the time step criterion for particle pusher they found would be too restrictive in the case of realistic pulsars or magnetars. In order to compare our results with their findings, we computed the invariant quantity (which is another expression of eq.~(\ref{eq:vitesse_onde_lineaire_0}))
\begin{equation}\label{eq:invariant}
\gamma \,  m_e \, c - p_{\rm x} = m_e \, c .
\end{equation}
We found that this invariant is well conserved for the full time of integration for any strength parameter giving always the value $1$ in normalized units. We show it however in a different manner compared to \cite{arefiev_temporal_2015}, instead of the above invariant, we plotted the Lorentz factors in Fig.~\ref{fig:onde_lineaire_gamma}. An estimate of the error is shown in fig.~\ref{fig:erreur_gamma_onde}. Here also our scheme is second order in time.

\subsection{Circularly polarized plane wave}

\begin{figure}
	\centering
	\includegraphics{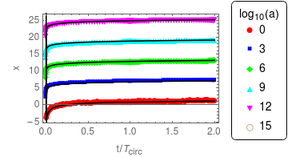}
	\caption{Motion of a positron in a circularly polarized plane wave for different strength parameters $a=10^i$ with $i\in\{0,3,6,9,12,15\}$. Symbols correspond to the analytical solution eq.~(\ref{eq:position_onde_circulaire}).}
	\label{fig:onde_circulaire}
\end{figure}

\begin{figure}
	\centering
	\includegraphics{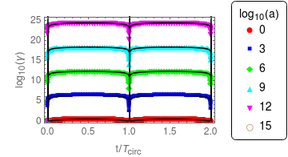}
	\caption{Lorentz factor of a positron in a circularly polarized plane wave for different strength parameters $a=10^i$ with $i\in\{0,3,6,9,12,15\}$. Symbols correspond to the analytical solution eq.~(\ref{eq:vitesse_onde_circulaire}).}
	\label{fig:onde_circulaire_gamma}
\end{figure}

Consider a circularly  polarized plane wave propagating in the $\ex$ direction such that the vector potential has components $A^\alpha=(0,0,\frac{E}{\omega}\,\cos\xi,\frac{E}{\omega}\,\sin\xi)$ and the wave vector $K^\alpha=(\frac{\omega}{c},k,0,0)$ thus the phase $\xi=\omega\,t-k\,x$. The electromagnetic field is then given by 
\begin{subequations}
	\begin{align}
	\mathbf{E} & = E \, ( \sin \xi \, \ey - \cos \xi \, \ez ) \\
	\mathbf{B} & = \frac{E}{c} \, ( \sin \xi \, \ez + \cos \xi \, \ey ) .
	\end{align}
\end{subequations}
Initially the particle is at rest with 4-velocity $u^\alpha_0=(c,\mathbf 0)$. The time evolution of the components of this 4-velocity will be
\begin{subequations}
	\label{eq:vitesse_onde_circulaire}
	\begin{align}
	u^x & = a^2 \, c \, (1 - \cos \xi ) = a \, u^y \\
	u^y & = a \, c \, (1 - \cos \xi ) \\
	u^z & = - a \, c \, \sin \xi \\
	u^0 & = c + u^x .
	\end{align}
\end{subequations}
The mean spatial velocity becomes 
\begin{subequations}
	\begin{align}
	<v^x> & = \frac{a^2\,c}{1+a^2} \\
	<v^y> & = \frac{a\,c}{1+a^2} \\
	<v^z> & = 0 .
	\end{align}
\end{subequations}
After integration, assuming the particle starts at rest at the origin at phase $\xi=0$, we find
\begin{subequations}
	\label{eq:position_onde_circulaire}
	\begin{align}
	\omega \, x & = a^2\,c \, (\xi - \sin\xi) \\
	\omega \, y & = a\,c\, (\xi - \sin\xi) \\
	\omega \, z & = a\,c\, (\cos\xi-1) \\
	\omega \, c \, t & = c\,\xi + \omega \, x .
	\end{align}
\end{subequations}
Examples of motion along the $x$ axis are shown in fig.~\ref{fig:onde_circulaire} for mildly and ultra-relativistic strength parameters $a=10^i$ with $i\in\{0,3,6,9,12,15\}$. The mean motion with average velocity $<v_x>$ is also shown. The corresponding evolution of the Lorentz factor is given in fig.~\ref{fig:onde_circulaire_gamma}. The numerical integration is compared to the analytical solution depicted by coloured symbols. Here again, both are in perfect agreement. We are even able to push the Lorentz factor limit well above $\gamma = \numprint{e15}$. Circular polarization is more efficient in accelerating particles to ultra-relativistic speeds.

\begin{figure}
	\centering
	\includegraphics[width=0.8\textwidth]{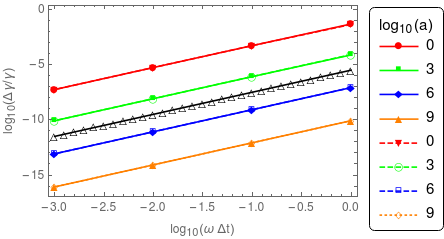}
	\caption{Relative error $\Delta\gamma/\gamma$ on the Lorentz factor at time $t=T/2$ corresponding to the maximum Lorentz factor reachable, depending on observer time step~$\Delta t$ on log-log scale for the linearly, solid line, and circularly, dashed line, polarized waves and strength parameter~$a$. The black solid line triangles correspond to a $\Delta t^{-2}$ slope.}
	\label{fig:erreur_gamma_onde}
\end{figure}

\begin{figure}
	\centering
	\includegraphics[width=0.8\textwidth]{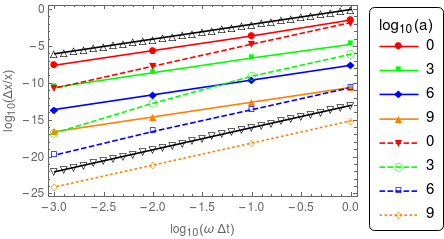}
	\caption{Relative error $\Delta x/x$ on the particle position at time $t=T/2$ corresponding to the maximum Lorentz factor reachable, depending on observer time step~$\Delta t$ on log-log scale for the linearly, solid line, and circularly, dashed line, polarized waves and strength parameter~$a$. The black solid line triangles correspond to a $\Delta t^{-2}$ slope and a $\Delta t^{-3}$ slope.}
	\label{fig:erreur_position_onde}
\end{figure}

The true motion in this plane wave is also perfectly periodic with a period given by
\begin{equation}\label{eq:periode_circulaire}
T_{\rm circ} = 2\,\pi\,\left( 1 + a^2 \right)
\end{equation}
the particle returning to rest at each period~$T_{\rm circ}$. The maximum Lorentz factor, reached at $t=T_{\rm circ}/2$ is the same as eq.~(\ref{eq:gamma_max}). Fig.~\ref{fig:erreur_gamma_onde} again show the error in the Lorentz factor for a circularly polarized wave, second order in time still holds. It overlaps with the linearly polarized wave and is indistinguishable.
The corresponding relative error in the $x$ position of the particle at $t=T_{\rm circ}/2$ is shown in Fig.~\ref{fig:erreur_position_onde} again for a linearly and a circularly polarized wave, second order in time still holds, we even observed an almost third order for the circularly polarized wave.
Note that for the highest strength parameters~$a$, in order to increase the precision of the computation, we tried a precision exceeding that of standard built-in types such as "double precision" codes written in C++. We used the boost multiprecision library with cpp\_dec\_float\_50 format with only modest additional computational cost less than a factor two and only small changes in the C++ code.

Eventually, we show an implementation of our new pusher in a 1D electromagnetic PIC code to demonstrate the feasibility of our approach, switching back to standard double precision built-in types. We emphasize however that for realistic neutron star applications, accuracy for large strength parameters is compulsory and therefore going beyond "double precision" must not be discarded. But in depth analysis of this requirement is left to another work as described by a companion paper (Tomczak \& P\'etri, submitted to JPP) specifically dealing with particle acceleration in ultra-strong electromagnetic fields of neutron stars.

\section{Simple 1D PIC tests}
\label{sec:PIC}

As an example of implementation of our new particle pusher, we developed a one dimensional relativistic PIC code with periodic boundary conditions for testing it on plasma oscillations and on the two-stream instability. For non periodic conditions, we tested it on a relativistic perpendicular shock problem.

In all our simulations, quantities are normalized according to some fundamental constants like the speed of light~$c$, the mass of one species of particles~$m_s$, its charge in absolute value~$|q_s|$ and a typical particle density number~$n_s$. From these constants we derive a typical plasma density
\begin{equation}\label{eq:plasma_frequency}
\omega_{\rm p_s}^2 = \frac{n_s\,q_s^2}{m_s \, \varepsilon_0}	
\end{equation}
the associated skin depth
\begin{equation}\label{eq:skin_depth}
\lambda_s = \frac{c}{\omega_{\rm p_s}}
\end{equation}
and a timescale $T_s=\omega_{\rm p_s}^{-1}$. The electric field is in units of
\begin{equation}\label{eq:EO}
E_0 = \sqrt{\frac{n_s\,m_s\,c^2}{\varepsilon_0}}
\end{equation}
and the magnetic field in units of~$B_0=E_0/c$.

In the subsequent simulation tests, we used an electron/ion plasma for oscillations but an electron/positron pair plasma with equal density number for the two-stream and shock runs.

\subsection{Plasma oscillations}

When a charge density~$\rho_{\rm s}=n_{\rm s}\,q_{\rm s}$ grows in a plasma, it tries to relax to its equilibrium state by building a restoring force induced by the electric field. Combined with its inertia, the plasma enters into an oscillatory motion described by plasma oscillations at the relativistic plasma frequency
\begin{equation}
\label{eq:relativistic_plasma_frequency}
\omega_{\rm p_{\rm s}}^2 = \frac{n_{\rm s}\,q_{\rm s}^2}{\left<\Gamma_{\rm s}\right>\,m_{\rm s}\,\varepsilon_0} .
\end{equation}
$\left<\Gamma_{\rm s}\right>$ is the average bulk Lorentz factor of the moving plasma slab. For a cold plasma, no wave can propagate but for a warm plasma, propagation is permitted in the form of Langmuir waves.

Plasma oscillations are purely electrostatic problems reducing Maxwell equations to the Maxwell-Gauss part
\begin{equation}\label{eq:Gauss}
\nabla \cdot \mathbf{E} = \frac{\rho_{\rm s}}{\varepsilon_0} .
\end{equation}
In our electrostatic PIC version of the code, we integrate this equation by fast Fourier transform techniques assuming periodic boundary conditions in the $z$ direction. In order to initiate plasma oscillations, we periodically perturb the initial position of the electrons evolving in an otherwise immobile net of ions. The motion becomes relativistic if the electric field energy density $\varepsilon_0\,E^2/2$ induced by the perturbation becomes equal to the rest mass energy density $n_{\rm s}\,m_{\rm s}\,c^2$.

\begin{figure}
	\centering
	\includegraphics[width=0.9\linewidth]{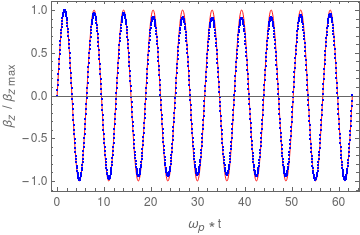}
	\caption{Non relativistic plasma oscillations showing the longitudinal velocity $\beta_{\rm z}$ normalized to its maximum value $\beta_{\rm zmax} \ll 1$, simulations in blue points and theory in red solid line. }
	\label{fig:oscillations1}
\end{figure}

\begin{figure}
	\centering
	\includegraphics[width=0.9\linewidth]{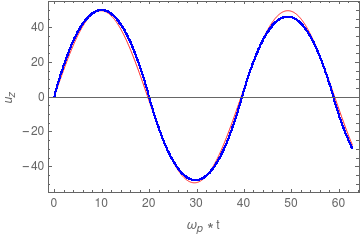}
	\caption{Relativistic plasma oscillations showing the longitudinal 4-velocity $u_{\rm z}$ with a mean factor $\left<\Gamma_{\rm s}\right> \approx 31$, simulations in blue points and fit in red solid line. }
	\label{fig:oscillations2}
\end{figure}

In figure~\ref{fig:oscillations1}, an example of non relativistic cold plasma oscillations shows the good agreement between the expectation and the simulations. A second example of a relativistic cold plasma oscillations is shown in figure~\ref{fig:oscillations2}. Here also, the simulations agree with equation~(\ref{eq:relativistic_plasma_frequency}). The temporal evolution of the 4-velocity $u_{\rm z}$ is fitted with a sinusoidal law
\begin{equation}\label{eq:uz}
 u_{\rm z} = A \, u_{\rm zmax} \, \sin(\omega\,t) .
\end{equation}
The corresponding average bulk Lorentz factor is $\left<\Gamma_{\rm s}\right> \approx 31$ whereas its maximum is $\left<\Gamma_{\rm s}\right>_{\rm max} \approx 50$.

\subsection{Two-stream instability}

\begin{figure}
	\centering
	\includegraphics[width=0.9\linewidth]{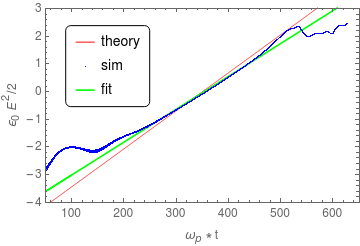}
	\caption{Relativistic two-stream instability with $\Gamma_{\rm b}=10$, simulations in blue points, theory in red solid line and fit in green. }
	\label{fig:instabilite}
\end{figure}

When two beams move through each other, an instability known as the two-stream instability grows, tapping some energy from the particle kinetic energy into the electric field. This instability is ubiquitous when beams of different bulk velocities penetrate each other. The dispersion relation for two counter-propagating streams of bulk velocity $v_{\rm b}$ and $-v_{\rm b}$ is
\begin{equation}\label{eq:twostream1}
 \frac{\omega_{\rm p_-}^2}{\gamma_{\rm b}^3\,(\omega-k\,v_{\rm b})^2} + \frac{\omega_{\rm p_+}^2}{\gamma_{\rm b}^3\,(\omega+k\,v_{\rm b})^2} = 1
\end{equation}
with $\omega$ the wave frequency and $k$ its wave number. The solution is given for an electron-positron plasma with $\omega_{\rm p_-}=\omega_{\rm p_+}=\omega_{\rm p}$ and $X=\gamma_{\rm b}^3 \, \frac{k^2\,v_{\rm b}^2}{\omega_{\rm p}^2}>0$ by
\begin{equation}\label{eq:twostream2}
\gamma_{\rm b}^3 \, \frac{\omega^2}{\omega_{\rm p}^2} = 1 + X \pm \sqrt{1+4\,X} .
\end{equation}
The counter-streaming plasma is unstable whenever $\omega^2<0$ corresponding to $X<2$. The minimum is reached at $X=3/4$ and it equals $-1/4$. The maximum growth rate $\omega_{\rm i} = \textrm{Im}(\omega)$ and its corresponding wave number are
\begin{equation}\label{eq:modemax}
 \frac{\omega_{\rm i}}{\omega_{\rm p}} = \frac{1}{2 \, \gamma_{\rm b}^{3/2}} \qquad ; \qquad \frac{k\,c}{\omega_{\rm p}} = \frac{\sqrt{3}}{2 \, \gamma_{\rm b}^{3/2}\,v_{\rm b}/c} . 
\end{equation}
An example of relativistic two-stream instability is shown in figure~\ref{fig:instabilite} for relativistic counter streaming beams with Lorentz factor each of $\Gamma_{\rm b}=10$. Simulations results show the total electrostatic energy $\varepsilon_0\,E^2/2$ in blue, compared with the expected maximum growth rate in red and a fit of the growth rate in green. The agreement between theory and simulations is good but some discrepancies are observed because our periodic simulation box is not exactly a multiple of the fastest growing wavelength. Therefore the simulation does not pick up rigorously the mode in equation~(\ref{eq:modemax}).

\subsection{Shock}

Finally for a case evolving the full electromagnetic field equations, we study a relativistic shock in 1D by colliding an electron/positron pair plasma onto a wall at rest in the simulation box.

Our simple 1D electromagnetic PIC code follows the ideas exposed in \cite{birdsall_plasma_2005} to solve the Maxwell equations by choosing the grid spacing~$\Delta x$ equal to the time step~$\Delta t$ times the speed of light $\Delta x = c\,\Delta t$. This technique has proven very useful to study strongly magnetised and relativistic shocks in pulsar winds as already done by \cite{lyubarsky_termination_2005} and \cite{petri_magnetic_2007}.

The initial flow is set up with an electric field $\mathbf{E} = E \, \ey$ and a magnetic field $\mathbf{B} = B \, \ez$  with $c\,B>E>0$ inducing an electric drift in the positive $\ex$ direction. Particles are reflected on a solid wall located at $x=L$ and there is no incoming flux of particles from the left $x=0$. The simulation frame therefore corresponds to the downstream plasma frame. In this frame, the upstream flow has a velocity $\mathbf{V}=E/B\,\ex$ with Lorentz factor $\Gamma=(1-V/c)^{-1/2}$. To compute the time step in the drifting frame, we assume that particles have constant velocity during the integration. For a good guess we can use eq.~(\ref{eq:DeltaTprime}) for the relation between observer frame~$\Delta t$ and drifting frame~$\Delta t'$ time steps. More sophisticated methods could release this assumption but we will show that this approximation already gives accurate results.

The magnetization parameter in the upstream flow defined (within a factor unity) by the ratio between the magnetic energy density over the plasma energy density is given by
\begin{equation}\label{eq:magnetization}
\sigma = \frac{B^2}{\mu_0\,\Gamma\,n\,m\,c^2} .
\end{equation}
A strongly magnetized flow implies $\sigma \gg 1$. In this limit, the downstream plasma density, temperature and Lorentz factor is summarized in table A.1 of \cite{petri_magnetic_2007}. As shown in this paper, for strong magnetizations and relativistic shock speeds, the compression ratio between the downstream plasma density~$n_2'$ and the upstream plasma density~$n_{1/2}$ as measured in the downstream plasma is
\begin{equation}\label{eq:density_jump}
\frac{n_2'}{n_{1/2}} = \frac{1+\beta_2}{\beta_2} \approx 2.
\end{equation}
Therefore the compression ratio is almost~2 if the downstream flow remains ultra relativistic with $\beta_2\approx1$.

\begin{figure}
	\centering
	\includegraphics[width=0.9\linewidth]{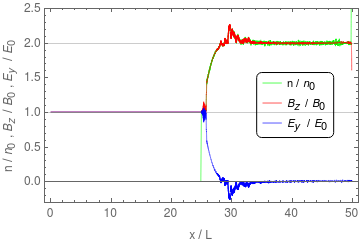}
	\caption{Upstream and downstream electromagnetic field $(E_y,B_z)$ and particle density number $n$ as a function of position~$x$ for an ultra-relativistic strongly magnetized MHD shock with $\Gamma\approx70$ and $\sigma\approx70$. All quantities are normalized with respect to their upstream value.}
	\label{fig:choc}
\end{figure}

A example of such a relativistic shock is shown in figure~\ref{fig:choc} for the upstream and downstream electromagnetic field $(E_y,B_z)$ and the particle density number~$n$, all normalized to their respective values. As a guide to the eye, the line $y=1$ and $y=2$ are also shown in grey. The flow bulk Lorentz factor is $\Gamma\approx70$ and the magnetization is $\sigma\approx70$. In this shock, the MHD jump conditions are satisfied with the magnetic flux conservation property stipulating that $B/n$ is conserved. This is indeed observed in the simulations. Inspecting  figure~\ref{fig:choc}, the compression ratio is indeed 2 as expected from an ultra-relativistic strongly magnetized MHD shock. The downstream electric field vanishes because the electric drift must disappear in this plasma, forced to be at rest by the solid wall boundary condition. Such flow are met for instance in pulsar wind nebulae where $\Gamma\gg1$ and $\sigma\gg1$ before entering the termination shock \citep{kirk_theory_2009}.
The simulation has been stopped when the back propagating electromagnetic wave has reach the middle of the simulation box. This is because there is no incoming particle flow from the left at $x=0$, therefore we observe a sharp jump in the density at $x=L/2$ from $n/n_0=1$ to $n/n_0=0$.

\section{Conclusions}
\label{sec:Conclusion}

We designed a new scheme for particle trajectory integration in any electromagnetic field configuration by analytically solving the relativistic equation of motion for a charged particle. The trajectory is given by an explicit closed analytical form free of any approximation as long as the field remains constant and uniform. For spatially and time dependent fields, numerical errors arise from the assumptions of constant fields when advancing to the next time step. Between two integration times, the motion must remains bound to size less than the typical space and time scales. These restrictions limit the size of the time step. Nevertheless, for plasmas with Larmor radii much smaller than the typical length scale of the electromagnetic field, this approximation must be excellent. It avoids resolving the gyro-period, enabling an increase by several orders of magnitude of the time step. Such approximations are particularly well suited for neutron star electromagnetic environments.

In a last part, we implemented our scheme in a fully relativistic electromagnetic 1D PIC code. We tested it against plasma oscillations, two-stream instabilities and strongly magnetized relativistic shocks, showing good accuracy. It demonstrated that our algorithm is viable for performing self-consistent plasma simulations. Nevertheless, a full 3D version of our PIC code would require more effort and was not the scope of our present paper. Nevertheless, such extensions are planed in the near future.

Moreover, when particles are accelerated to ultra-relativistic speeds, they usually radiate copiously photons that carry energy and momentum into the radiation field which feeds back to the particle equation of motion via radiation reaction. This effect is very important in limiting the maximum Lorentz factor reachable by charged particles. We plan to include this radiation reaction in our exact analytical integration scheme in a forthcoming paper.

\section*{Acknowledgements}

This work has been supported by CEFIPRA grant IFC/F5904-B/2018. We also acknowledge the High Performance Computing center of the University of Strasbourg for supporting this work by providing scientific support and access to computing resources. Part of the computing resources were funded by the Equipex Equip@Meso project (Programme Investissements d'Avenir) and the CPER Alsacalcul/Big Data.

\bibliographystyle{jpp}
\bibliography{/home/petri/zotero/Ma_bibliotheque} 

\end{document}